\documentclass[english,eqsecnum,nofootinbib,superscriptaddress]{revtex4-2}
\usepackage[T1]{fontenc}
\usepackage[utf8]{inputenc}
\usepackage{color}
\usepackage{colortbl}
\usepackage{babel}
\usepackage{float}
\usepackage{mathrsfs}
\usepackage{mathtools}
\usepackage{bm}
\usepackage{amsmath}
\usepackage{amssymb}
\usepackage[pdfusetitle,
 bookmarks=true,bookmarksnumbered=false,bookmarksopen=false,
 breaklinks=false,pdfborder={0 0 1},backref=false,colorlinks=true]
 {hyperref}

\makeatletter

\providecommand{\tabularnewline}{\\}

\usepackage{xcolor}
\usepackage{datetime2}
\usepackage{extarrows}

\makeatother

\begin{document}
\title{Degrees of freedom of a quadratic scalar-nonmetricity theory}
\author{Jia-Jun Chen}
\affiliation{School of Physics, Sun Yat-sen University, Guangzhou 510275, China}
\author{Zheng Chen}
\affiliation{School of Physics and Astronomy, Sun Yat-sen University, Zhuhai 519082,
China}
\author{Xian Gao}
\email[Corresponding author: ]{gaoxian@mail.sysu.edu.cn}

\affiliation{School of Physics, Sun Yat-sen University, Guangzhou 510275, China}
\affiliation{Guangdong Provincial Key Laboratory of Quantum Metrology and Sensing,
Sun Yat-sen University, Zhuhai 519082, China}
\begin{abstract}
We study the number of degrees of freedom (DOFs) in quadratic scalar-nonmetricity
(QSN) theory, whose Lagrangian is the linear combination of five quadratic
nonmetricity invariants with coefficients depending on a dynamical
scalar field. Working in the coincident gauge, we perform the Arnowitt-Deser-Misner
decomposition and classify QSN models into 13 cases according to the
numbers of their primary constraints. For cases that are physically
viable in the sense that both a consistent cosmological background
and tensor gravitational waves exist, we count the number of DOFs
based on two approaches. First, we investigate the linear cosmological
perturbations around a Friedmann-Lema\^{\i}tre-Robertson-Walker  background. Then we perform a Dirac-Bergmann
Hamiltonian constraint analysis to count the number of DOFs at the
nonperturbative level. We focus on three representative cases. In
case II, both the perturbative and nonperturbative approaches yield
the same result, which indicates that the theory propagates 10 degrees
of freedom. In contrast, in cases V and VI, the Hamiltonian analysis
yields 8 degrees of freedom, while only 6 and 5 modes are visible
at linear order in perturbations, respectively. This indicates that
additional modes are strongly coupled on cosmological backgrounds.
\end{abstract}
\maketitle

\section{Introduction \label{sec:intro}}

In recent years, there has been growing interest in reformulating
gravity using non-Riemannian geometric structures, in which the gravitational
interaction is attributed to torsion and/or nonmetricity rather than
curvature (see \citep{Heisenberg:2023lru} for a review). As a natural
extension of general relativity, the metric-affine gauge theory of
gravity provides a comprehensive framework that incorporates both
torsion and nonmetricity as independent dynamical fields \citep{Hehl:1994ue}.
In the symmetric teleparallel description, one employs a curvature-free
and torsionless connection and characterizes the geometry by the nonmetricity
tensor $Q_{\rho\mu\nu}\equiv\nabla_{\rho}g_{\mu\nu}$. Within this
framework, symmetric teleparallel gravity and its extensions [most
notably $f(Q)$ gravity] have been studied extensively \citep{Nester:1998mp,Adak:2005cd,Adak:2006rx,Adak:2008gd,Mol:2014ooa,BeltranJimenez:2018vdo,BeltranJimenez:2019tme,Lazkoz:2019sjl,Lu:2019hra,DAmbrosio:2020nqu,Xu:2020yeg,Zhao:2021zab,Albuquerque:2022eac,BeltranJimenez:2022azb,Dimakis:2022rkd}. 

A useful perspective is provided by the ``geometric trinity,'' according
to which curvature, torsion, and nonmetricity offer three equivalent
and complementary descriptions of general relativity (GR). In particular,
these equivalent formulations propagate the same number of degrees
of freedom (DOFs) as GR \citep{Blagojevic:2000pi,Blagojevic:2000qs,DAmbrosio:2020nqu,Maluf:2001rg}.
However, once one goes beyond GR to nonlinear modifications such as
$f(Q)$ and $f(T)$, this equivalence is generally lost. The associated
DOF content becomes subtle and model-dependent, which has motivated
substantial discussion \citep{Li:2011rn,Ong:2013qja,Ferraro:2016wht,Ferraro:2018tpu,Hu:2022anq,Blagojevic:2020dyq,Dimakis:2022rkd,Tomonari:2023wcs,DAmbrosio:2023asf,Gomes:2023tur}.
A key reason is that the boundary term relating the different geometric
scalars is itself modified in the nonlinear extensions, and therefore
cannot be ignored in the action. In this sense, theories incorporating
the boundary term explicitly [e.g., $f(Q,B)$ or $f(T,B)$] provide
a natural arena to track how the trinity relation is altered. More
broadly, these non-Riemannian modified gravities can introduce additional
propagating DOFs associated with torsion and/or nonmetricity, leading
to rich phenomenology in cosmology \citep{BeltranJimenez:2019tme,Lu:2019hra,Xu:2020yeg,Albuquerque:2022eac},
black hole solutions \citep{Wang:2021zaz,Lin:2021uqa,DAmbrosio:2021zpm,Bahamonde:2022esv,Calza:2022mwt},
and gravitational waves \citep{Hohmann:2018wxu,Hohmann:2018xnb,Soudi:2018dhv,Conroy:2019ibo,Chatzistavrakidis:2020wum,Cai:2021uup,Chen:2022wtz}. 

Another well-motivated direction is to consider general couplings
between a scalar field and torsion and/or nonmetricity, i.e., “scalar-torsion”
and “scalar-nonmetricity” theories \citep{Bahamonde:2017wwk,Hohmann:2018xnb,Jarv:2018bgs,Runkla:2018xrv,Soudi:2018dhv,Bahamonde:2019shr}.
When the scalar field possesses a spacelike gradient (as in cosmological
settings), the scalar-tensor theories in the Riemannian framework
can be recast as metric theories with only spatial covariance, leading
to a framework  dubbed spatially covariant gravity \citep{Gao:2014soa,Gao:2014fra,Gao:2020yzr,Hu:2021bbo,Hu:2021yaq}.
This approach has been generalized to scalar-nonmetricity theories,
leading to spatially covariant nonmetricity gravity \citep{Yu:2024drx}.
Parity-violating extensions have also been explored in both torsional
and nonmetric formulations \citep{Conroy:2019ibo,Chatzistavrakidis:2020wum,Li:2020xjt,Iosifidis:2020dck,Cai:2021uup,Wu:2021ndf,Zhao:2021zab,Iosifidis:2021bad,Rao:2021azn,Li:2021mdp,Li:2022mti,Li:2022vtn,Chen:2022wtz}. 

In this work we study the number of DOFs of a class of scalar-nonmetricity
models whose Lagrangian is the most general linear combination of
the five independent quadratic invariants constructed from $Q_{\rho\mu\nu}$,
with coefficients promoted to arbitrary functions of a dynamical scalar
field. We refer to this class of theories as quadratic scalar-nonmetricity
(QSN) theory. This setup is a straightforward scalar extension of
``Newer GR'' theory proposed in \citep{BeltranJimenez:2017tkd}.
The newer GR theory has been extensively studied \citep{Hohmann:2018wxu,Flathmann:2020zyj,Golovnev:2024owe,Hohmann:2024phz,Hohmann:2025kxp}.
When the coefficient functions become constants, one recovers the
purely geometric quadratic nonmetricity theories. The primary constraint
structure was previously analyzed via the nonmetricity components
and the Hessian matrix in \citep{Dambrosio:2020wbi}, and related
issues in general teleparallel quadratic gravity were studied in \citep{Bajardi:2024qbi}.
Recently, the number of DOFs of Newer GR has also been studied \citep{Ganz:2025ydt}.
For our purposes, a particularly transparent approach is to perform
an Arnowitt-Deser-Misner (ADM) decomposition (in a convenient gauge)
and examine directly which velocities can be solved in terms of canonical
momenta, thereby obtaining explicit primary constraints and enabling
a systematic analysis of their evolution. To our knowledge, the number
of DOFs in this general QSN class has not been studied systematically
before, even in the case with constant coefficients. Thus, providing
such a classification is one of the main motivations of the current
work. 

A further motivation comes from the strong coupling issues identified
in teleparallel/nonmetric modifications. It is by now well appreciated
that $f(Q)$ and $f(T)$ theories can exhibit strong coupling on cosmological
backgrounds \citep{Gomes:2023tur,Hu:2023xcf,Hu:2023juh}. This raises
a natural question: do analogous strong coupling phenomena arise in
scalar-nonmetricity (and scalar-torsion) theories? Addressing this
requires comparing the DOFs inferred from linear perturbations with
those obtained from a fully nonperturbative Hamiltonian constraint
analysis, since modes that are absent at linear order may reappear
nonlinearly as strongly coupled degrees of freedom. 

Concretely, we work in the coincident gauge and perform the ADM decomposition
of the QSN action, and then classify QSN models into distinct cases according
to the number of primary constraints. For cases that are phenomenologically
viable, in the sense that they admit a consistent Friedmann-Lema\^{\i}tre-Robertson-Walker (FLRW) background
and support tensor gravitational waves, we count DOFs using two complementary
methods. One is the linear cosmological perturbations around FLRW
and the other is a Dirac-Bergmann Hamiltonian analysis at the nonperturbative
level. We focus on three representative cases. In case II, both approaches
agree and show that the theory propagates 10 DOFs. In contrast, in
cases V and VI the Hamiltonian analysis yields 8 DOFs, whereas only
6 and 5 modes, respectively, are visible at linear order in cosmological
perturbations. This mismatch indicates that additional modes are strongly
coupled on cosmological backgrounds.

This paper is organized as follows. In Sec. \ref{sec:QSN} we introduce
the QSN theory. In Sec. \ref{sec:pricons} we derive its ADM decomposition
in the coincident gauge, obtain the conjugate momenta, and present
a systematic classification by primary constraints. In Sec. \ref{sec:pert}
we analyze linear cosmological perturbations and determine the DOFs
visible at quadratic order for the cases of interest. In Sec. \ref{sec:sec_cons},
we perform the Dirac-Bergmann Hamiltonian constraint analysis and
compute the nonperturbative DOF counts, highlighting when and how
strong coupling arises. We work in units where $8\pi G=1$ and use
the metric signature $\{-,+,+,+\}$. 

\section{Quadratic scalar-nonmetricity theory \label{sec:QSN}}

The nonmetricity tensor is defined by
\begin{equation}
Q_{\rho\mu\nu}\coloneqq\nabla_{\rho}g_{\mu\nu},
\end{equation}
where $g_{\mu\nu}$ is the spacetime metric and $\nabla$ is a general
affine connection\footnote{We take the convention such that for a vector field $V^{\mu}$, $\nabla_{\mu}V^{\nu}=\varGamma^{\nu}_{\phantom{\nu}\rho\mu}V^{\rho}$.}.
For later convenience, we denote 
\begin{equation}
Q_{\mu}\equiv Q^{\phantom{\mu}\rho}_{\mu\phantom{\rho}\rho}=g^{\rho\sigma}Q_{\mu\rho\sigma},\quad q_{\mu}\equiv Q^{\rho}_{\phantom{\rho}\rho\mu}=g^{\rho\sigma}Q_{\rho\sigma\mu},
\end{equation}
as shorthand. As in symmetric teleparallelism, we assume that the
affine connection is free of curvature and torsion, i.e.,
\begin{equation}
R^{\mu}_{\phantom{\mu}\nu\rho\sigma}\equiv\partial_{\rho}\varGamma^{\mu}_{\phantom{\mu}\nu\sigma}-\partial_{\sigma}\varGamma^{\mu}_{\phantom{\mu}\nu\rho}+\varGamma^{\mu}_{\phantom{\mu}\lambda\rho}\varGamma^{\lambda}_{\phantom{\lambda}\nu\sigma}-\varGamma^{\mu}_{\phantom{\mu}\lambda\sigma}\varGamma^{\lambda}_{\phantom{\lambda}\nu\rho}=0
\end{equation}
and
\begin{equation}
T^{\rho}_{\phantom{\rho}\mu\nu}\equiv\varGamma^{\rho}_{\phantom{\rho}\nu\mu}-\varGamma^{\rho}_{\phantom{\rho}\mu\nu}=0.
\end{equation}
As a result, the coefficients of the affine connection take the general
form \citep{Adak:2008gd,BeltranJimenez:2017tkd}
\begin{equation}
\varGamma^{\alpha}_{\phantom{\alpha}\beta\mu}=\frac{\partial x^{\alpha}}{\partial\xi^{a}}\frac{\partial^{2}\xi^{a}}{\partial x^{\mu}\partial x^{\beta}},\label{xi_def}
\end{equation}
where $\xi^{a}=\xi^{a}\left(x\right)$ with $a=0,1,2,3$ are four
general scalar fields. It is well-known that in the presence of the
nonmetricity tensor, we have the relation
\begin{equation}
\mathring{R}+Q+\mathring{\nabla}_{\mu}\left(Q^{\mu}-q^{\mu}\right)=0,
\end{equation}
where $\mathring{R}$ and $\mathring{\nabla}$ are the Ricci scalar
and the covariant derivative adapted to the metric-compatible Levi-Civita
connection, and $Q$ is the nonmetricity scalar defined by\footnote{Some authors define the nonmetricity scalar with an opposite sign
such that $\mathbb{Q}\equiv-Q\simeq\mathring{R}$.}
\begin{equation}
Q\coloneqq\frac{1}{4}Q_{\rho\mu\nu}Q^{\rho\mu\nu}-\frac{1}{2}Q^{\rho\mu\nu}Q_{\mu\nu\rho}-\frac{1}{4}Q_{\mu}Q^{\mu}+\frac{1}{2}q^{\mu}Q_{\mu}.
\end{equation}
In this sense, the Lagrangian $-Q\simeq\mathring{R}$ provides an
equivalent description of general relativity, which is dubbed the
symmetric teleparallel equivalent of general relativity (STEGR) \citep{Nester:1998mp}.

In general, there are five independent scalars quadratic in the nonmetricity
tensor, leading to the combination\footnote{The five quadratic nonmetricity invariants can be viewed as the parity-even
quadratic building blocks familiar from the metric-affine gauge-theory
literature. See, e.g., \citep{Hehl:1994ue} for the general irreducible
decomposition framework.}
\begin{equation}
\mathscr{Q}=c_{1}Q_{\rho\mu\nu}Q^{\rho\mu\nu}+c_{2}Q^{\rho\mu\nu}Q_{\mu\nu\rho}+c_{3}Q_{\mu}Q^{\mu}+c_{4}q_{\mu}q^{\mu}+c_{5}Q_{\mu}q^{\mu},
\end{equation}
where $c_{i}$ are arbitrary real numbers. The STEGR is recovered
by choosing $c_{1}=1/4$, $c_{2}=-1/2$, $c_{3}=-1/4$, $c_{4}=0$,
and $c_{5}=1/2$. In this work, we consider their nonminimal couplings
with a scalar field by promoting $c_{i}$ to  arbitrary functions
of $\phi$. The action is given by
\begin{align}
S_{\mathrm{QSN}}= & \int\mathrm{d}^{4}x\sqrt{-g}\Big[c_{1}\left(\phi\right)Q_{\rho\mu\nu}Q^{\rho\mu\nu}+c_{2}\left(\phi\right)Q^{\rho\mu\nu}Q_{\mu\nu\rho}+c_{3}\left(\phi\right)Q_{\mu}Q^{\mu}+c_{4}\left(\phi\right)q_{\mu}q^{\mu}+c_{5}\left(\phi\right)Q_{\mu}q^{\mu}\nonumber \\
 & \qquad-\frac{1}{2}\nabla_{\mu}\phi\nabla^{\mu}\phi+V\left(\phi\right)\Big],\label{QSN_action}
\end{align}
where the scalar field $\phi$ acquires its own kinetic term. The
scalar field $\phi$ thus plays the role of the matter content of
the universe. We will use (\ref{QSN_action}) as our starting point
to count the number of degrees of freedom of quadratic scalar-nonmetricity
(QSN) theory.

At this point, note that the widely studied $f(Q)$ theory is also
included in (\ref{QSN_action}) as a special case, since \citep{Jarv:2018bgs,Hu:2022anq}
\begin{equation}
S_{f(Q)}=\int\mathrm{d}^{4}x\sqrt{-g}f\left(Q\right)=\int\mathrm{d}^{4}x\sqrt{-g}\left[f'\left(\phi\right)Q+f\left(\phi\right)-\phi f'\left(\phi\right)\right].\label{SfQ}
\end{equation}
Therefore, $f(Q)$ theory is recovered by choosing $4c_{1}=-2c_{2}=-4c_{3}=2c_{5}=f'(\phi)$,
$c_{4}=0$, $V(\phi)=f\left(\phi\right)-\phi f'\left(\phi\right)$,
and turning off the kinetic term of the scalar field. The number of
DOFs of $f(Q)$ theory has been debated in the literature \citep{DAmbrosio:2020nqu,Tomonari:2023wcs,Heisenberg:2025fxc,Paliathanasis:2023pqp,Gomes:2023tur,Guzman:2023oyl,Zhao:2024kri,Dambrosio:2020wbi,DAmbrosio:2023asf,Heisenberg:2023wgk,Hu:2022anq,Hu:2023gui}.
We will come back to this issue in Sec. \ref{subsec:fQ-theory}.

\section{ADM formalism and primary constraints \label{sec:pricons}}

To perform a constraint analysis in the Hamiltonian formalism, we
first make a spacetime decomposition of the QSN action (\ref{QSN_action})
by employing the standard ADM formalism. To
simplify the calculations, in the following we work in the so-called
``coincident gauge'' by fixing $\xi^{a}=0$ and thus $\varGamma^{\alpha}_{\phantom{\alpha}\beta\mu}=0$
\citep{BeltranJimenez:2017tkd}. In the coincident gauge, $Q_{\rho\mu\nu}\equiv\partial_{\rho}g_{\mu\nu}$,
and thus the variables in the gravity sector are just the metric variables. 

We choose the standard ADM variables for the metric defined by
\begin{equation}
g_{\mu\nu}=\left(\begin{array}{cc}
-N^{2}+N_{i}N^{i} & N_{j}\\
N_{i} & h_{ij}
\end{array}\right),\label{eq:adm metric}
\end{equation}
where $i,j$ are the spatial indices. After tedious calculations,
we obtain the ADM decomposition of QSN action in the coincident gauge
\begin{align}
S_{\mathrm{QSN}}= & \int\mathrm{d}t\mathrm{d}^{3}x\sqrt{h}\bigg\{ N\left(\overset{3}{\mathcal{Q}}-4c_{1}K^{ij}K_{ij}-4c_{3}K^{2}\right)+\left(2c_{1}+c_{2}+c_{4}\right)\frac{h_{ij}\dot{N}^{i}\dot{N}^{j}}{N^{3}}-4\left(c_{1}+c_{2}+c_{3}+c_{4}+c_{5}\right)\frac{\dot{N}^{2}}{N^{3}}\nonumber \\
 & -4\left(2c_{3}+c_{5}\right)\frac{K\dot{N}}{N}-2\left(4c_{3}+c_{5}\right)K\partial_{i}N^{i}+4\left(2c_{3}+c_{5}\right)\frac{KN^{i}\partial_{i}N}{N}-4\left(2c_{1}+c_{2}\right)h^{ij}K_{kj}\partial_{i}N^{k}\nonumber \\
 & -2\left(4c_{3}+2c_{4}+3c_{5}\right)\dot{N}\frac{\partial_{i}N^{i}}{N^{2}}+8\left(c_{1}+c_{2}+c_{3}+c_{4}+c_{5}\right)\frac{N^{i}\dot{N}\partial_{i}N}{N^{3}}\nonumber \\
 & +\left[-2\left(2c_{1}+c_{2}+c_{4}\right)\frac{h_{ik}N^{j}\partial_{j}N^{i}}{N^{3}}-2c_{4}\frac{h^{ij}Q_{jki}}{N}-\frac{c_{5}h^{ij}Q_{kij}}{N}-2\left(2c_{2}+c_{5}\right)\frac{\partial_{k}N}{N^{2}}\right]\dot{N}^{k}\nonumber \\
 & +2h^{ij}h^{kl}\left(2c_{3}Q_{jkl}+c_{5}Q_{ljk}\right)\partial_{i}N+2\left(4c_{3}+2c_{4}+3c_{5}\right)\frac{\partial_{i}N^{i}N^{j}\partial_{j}N}{N^{2}}+2\left(2c_{2}+c_{5}\right)\frac{N^{i}\partial_{i}N^{j}\partial_{j}N}{N^{2}}\nonumber \\
 & +\partial_{i}N^{j}\left(-\frac{\left(2c_{1}+3c_{2}\right)\partial_{j}N^{i}}{N}+c_{5}\frac{h^{mn}N^{i}Q_{jmn}}{N}+2c_{4}\frac{h^{mn}N^{i}\partial_{n}h_{lm}}{N}+\frac{\left(2c_{1}+c_{2}+c_{4}\right)h_{lj}N^{k}N^{i}\partial_{k}N^{l}}{N^{3}}\right)\nonumber \\
 & -4\left(c_{1}+c_{2}+c_{3}+c_{4}+c_{5}\right)\frac{N^{i}N^{j}\partial_{i}N\partial_{j}N}{N^{3}}+4\left(c_{1}+c_{3}\right)\frac{h^{ij}\partial_{i}N\partial_{j}N}{N}-\left(4c_{3}+c_{4}+2c_{5}\right)\frac{\partial_{i}N^{i}\partial_{j}N^{j}}{N}\nonumber \\
 & -2\left(2c_{1}+c_{2}\right)\frac{h_{ij}h^{kl}\partial_{k}N^{i}\partial_{l}N^{j}}{N}+\frac{\dot{\phi}^{2}}{2N}-\frac{N^{i}\partial_{i}\phi\dot{\phi}}{N}-\frac{1}{2}Nh^{ij}\partial_{i}\phi\partial_{j}\phi+\frac{1}{2}\frac{N^{i}N^{j}\partial_{i}\phi\partial_{j}\phi}{N}+NV\left(\phi\right)\bigg\},\label{action_ADM}
\end{align}
where $N$ is the lapse function, $N^{i}$ is the shift vector, $h_{ij}$
is the spatial metric, and $h^{ij}$ is its inverse. In the above,
we define the three-dimensional nonmetricity scalar 
\begin{equation}
\overset{3}{\mathcal{Q}}\coloneqq\left(c_{1}h^{il}h^{jm}h^{kn}+c_{2}h^{in}h^{jm}h^{kl}+c_{3}h^{il}h^{jk}h^{mn}+c_{4}h^{ij}h^{lm}h^{kn}+c_{5}h^{ij}h^{mn}h^{kl}\right)Q_{ijk}Q_{lmn}.
\end{equation}
The details of deriving (\ref{action_ADM}) can be found in Appendix
\ref{app:decnm}. 

The connection in symmetric teleparallel geometry (\ref{xi_def})
can be written in terms of four St\"{u}ckelberg fields $\xi^{a}(x)$,
and the coincident gauge corresponds to the gauge choice $\xi^{a}=0$
(hence $\Gamma^{\alpha}{}_{\beta\mu}=0$). In this work we impose
this gauge at the level of the action to simplify the ADM decomposition.
Since this gauge choice fixes the coordinate freedom associated with
$\xi^{a}$, the resulting gauge-fixed action (\ref{action_ADM}) is
not manifestly diffeomorphism invariant\footnote{This can be seen explicitly from the appearance of terms such as $\partial_{j}N^{i}$,
which are clearly not spatially covariant.}. As a result, the usual split into first-/second-class constraints
is correspondingly reshuffled. Nevertheless, the number of physical
DOFs is expected to be gauge independent. One may equivalently start
from the covariant formulation with $\xi^{a}$ kept, perform the Dirac-Bergmann
analysis, and only then impose $\xi^{a}=0$ as gauge-fixing conditions,
which does not change the physical DOF count. In the following, our
classification is based on which velocities in $\{N,N^{i},h_{ij},\phi\}$
can be solved from the canonical momenta in the gauge-fixed formulation,
leading to explicit primary constraints listed in Appendix \ref{app:pri_cons}.

By choosing the special coefficients $c_{1}=1/4$, $c_{2}=-1/2$,
$c_{3}=-1/4$, $c_{4}=0$, $c_{5}=1/2$, and suppressing the kinetic
term of the scalar field, the action (\ref{action_ADM}) reduces to
the STEGR case \citep{DAmbrosio:2020nqu}
\begin{align}
S_{\mathrm{STEGR}}= & \int\mathrm{d}t\mathrm{d}^{3}x\sqrt{h}\bigg\{ N\left(\overset{3}{\mathcal{Q}}+K^{ij}K_{ij}-K^{2}\right)-K\partial_{i}N^{i}\nonumber \\
 & +\dot{N}\left(\frac{\partial_{i}N^{i}}{N^{2}}\right)+\left[\frac{1}{2N^{2}}\left(Nh^{ij}Q_{kij}-\partial_{k}N\right)\right]\dot{N}^{k}\nonumber \\
 & +h^{ij}h^{kl}\left(Q_{jkl}-Q_{ljk}\right)\partial_{i}N-\frac{\partial_{i}N^{i}N^{j}\partial_{j}N}{N^{2}}\nonumber \\
 & +\frac{N^{i}\partial_{i}N^{j}\partial_{j}N}{N^{2}}-\frac{\partial_{i}N^{j}}{2N}\left(2\partial_{j}N^{i}+h^{mn}N^{i}Q_{jmn}\right).
\end{align}

The ADM variables in QSN theory are $\Phi^{I}\equiv\{\phi,N,N^{i},h_{ij}\}$
with conjugate momenta denoted by $\Pi_{I}\equiv\{\pi_{\phi},\pi_{N},\pi_{i},\pi^{ij}\}$.
The canonical momenta are calculated as
\begin{equation}
\Pi_{I}\coloneqq\frac{\delta S}{\delta\dot{\Phi}^{I}},
\end{equation}
which are given explicitly by
\begin{equation}
\pi_{\phi}=\sqrt{h}\left(\frac{\dot{\phi}}{N}-\frac{N^{i}\partial_{i}\phi}{N}\right),\label{pi_phi}
\end{equation}
\begin{align}
\pi_{N}= & \sqrt{h}\bigg[-8\left(c_{1}+c_{2}+c_{3}+c_{4}+c_{5}\right)\frac{\dot{N}}{N^{3}}-4\left(2c_{3}+c_{5}\right)\frac{K}{N}-2\left(4c_{3}+2c_{4}+3c_{5}\right)\frac{\partial_{i}N^{i}}{N^{2}}\nonumber \\
 & \quad+8\left(c_{1}+c_{2}+c_{3}+c_{4}+c_{5}\right)\frac{N^{i}\partial_{i}N}{N^{3}}\bigg],\label{pi_N}
\end{align}
\begin{align}
\pi_{i}= & \sqrt{h}\bigg[-2\left(2c_{1}+c_{2}+c_{4}\right)\frac{h_{ki}N^{j}\partial_{j}N^{k}}{N^{3}}-2c_{4}\frac{h^{kj}Q_{jik}}{N}-\frac{c_{5}h^{kj}Q_{ikj}}{N}\nonumber \\
 & -2\left(2c_{2}+c_{5}\right)\frac{\partial_{i}N}{N^{2}}+2\left(2c_{1}+c_{2}+c_{4}\right)\frac{h_{ij}\dot{N}^{j}}{N^{3}}\bigg],
\end{align}
and
\begin{align}
\pi^{ij}= & \sqrt{h}\bigg[-4c_{1}K^{ij}-4c_{3}Kh^{ij}-4\left(2c_{3}+c_{5}\right)\frac{\dot{N}}{2N^{2}}h^{ij}-4\left(2c_{3}+c_{5}\right)\frac{\partial_{i}N^{i}}{2N}h^{ij}\nonumber \\
 & +4\left(2c_{3}+c_{5}\right)\frac{N^{k}\partial_{k}N}{2N^{2}}h^{ij}-4\left(2c_{1}+c_{2}\right)\frac{\partial_{k}N^{i}}{2N}h^{kj}\bigg].\label{pi^ij}
\end{align}

With generic values of $c_{1},\ldots,c_{5}$, all 11 configuration
variables $\Phi^{I}\equiv\{\phi,N,N^{i},h_{ij}\}$ enter the gauge-fixed
action with time derivatives. So no primary constraints arise from
the definition of momenta. Our purpose in this work is to make a systematic
classification of all the possible cases of coefficients according
to the corresponding number of DOFs. To this end, we first need to
assume that the Lagrangian is degenerate; that is, some velocities
cannot be solved in terms of their conjugate momenta (together with
the variables), which implies the existence of primary constraints. 

We can tune the coefficients of velocities $\dot{\Phi}^{I}$ in (\ref{pi_phi})-(\ref{pi^ij})
to generate different numbers of primary constraints. We exhaust all
the possible cases, which are listed in Tables \ref{tab:cases_phy}
and \ref{tab:cases_unphy}. We find there are a total of 13
cases, which we divide into two classes according to whether $c_{1}$
vanishes or not. In the first class, which is shown in Table \ref{tab:cases_phy},
$c_{1}\neq0$ and the corresponding theories will contribute to the
tensor modes (i.e., gravitational waves in homogeneous and isotropic
background). The corresponding primary constraints in various cases
can be found in Appendix \ref{app:pri_cons}. According to (\ref{SfQ}),
the widely studied $f(Q)$ theory in the presence of the kinetic term
of the scalar field belongs to case VI. Note STEGR also belongs to
case VI, which can be understood as a special limit of $f(Q)\rightarrow Q$
with a dynamical scalar field. 
\begin{table}[H]
\begin{centering}
\begin{tabular}{|c|c|c|c|c|c|c|c|}
\hline 
Cases & I & II & III & IV & V & VI (GR) & VII\tabularnewline
\hline 
Conditions & - & $\begin{array}{c}
c_{4}=c_{3}-c_{1}-c_{2}\\
c_{5}=-2c_{3}
\end{array}$ & $\begin{array}{c}
c_{1}=-3c_{3}\\
c_{5}=-2c_{3}
\end{array}$ & $\begin{array}{c}
c_{1}=-3c_{3}\\
c_{5}=-2c_{3}\\
c_{4}=4c_{3}-c_{2}
\end{array}$ & $c_{4}=-2c_{1}-c_{2}$ & $\begin{array}{c}
c_{3}=-c_{1}\\
c_{4}=-2c_{1}-c_{2}\\
c_{5}=2c_{1}
\end{array}$ & $\begin{array}{c}
c_{4}=-2c_{1}-c_{2}\\
c_{1}=-3c_{3}\\
c_{5}=-2c_{3}
\end{array}$\tabularnewline
\hline 
No. of p.c. & 0 & 1 & 1 & 2 & 3 & 4 & 4\tabularnewline
\hline 
\end{tabular}
\par\end{centering}
\caption{Cases with GWs.}

\label{tab:cases_phy}
\end{table}
In the second class, which is shown in Table \ref{tab:cases_unphy},
all the cases share the same condition $c_{1}=0$, which indicates
the absence of tensor gravitational waves. This can also be seen explicitly
in the next section. 

\begin{table}[H]
\begin{centering}
\begin{tabular}{|c|c|c|c|c|c|c|}
\hline 
Cases & VIII & IX & X & XI & XII & XIII\tabularnewline
\hline 
Conditions & $c_{1}=0$ & $\begin{array}{c}
c_{1}=0\\
c_{4}=-c_{2}+c_{3}\\
c_{5}=-2c_{3}
\end{array}$ & $\begin{array}{c}
c_{1}=0\\
c_{5}=0\\
c_{3}=0
\end{array}$ & $\begin{array}{c}
c_{1}=0\\
c_{2}=-c_{4}
\end{array}$ & $\begin{array}{c}
c_{1}=0\\
c_{2}+c_{4}=0\\
c_{3}=-c_{5}
\end{array}$ & $\begin{array}{c}
c_{1}=0\\
c_{2}=-c_{4}\\
c_{3}=0\\
c_{5}=0
\end{array}$\tabularnewline
\hline 
No. of p.c. & 5 & 6 & 6 & 8 & 9 & 10\tabularnewline
\hline 
\end{tabular}
\par\end{centering}
\caption{Cases without GWs.}

\label{tab:cases_unphy}
\end{table}

Note the classification above has included the cases discussed in
\citep{Dambrosio:2020wbi}. Here we further classify them into two
classes according to whether there are tensor gravitational waves.
Compared to \citep{Dambrosio:2020wbi}, our classification differs
in that we allow $c_{i}$ to depend on a dynamical scalar $\phi$,
which can change the degeneracy conditions. In the limit of constant
$c_{i}$, our classification reduces to the known purely geometric
subclasses. Moreover, we diagnose degeneracy directly at the level
of the ADM momenta in the coincident gauge, which makes certain primary
constraints manifest. Precisely, in the present work, there is a case
corresponding to a single primary constraint (case II), and we also
have cases corresponding to 2 (case IV) and 3 (case V) primary constraints.
In addition, in case XII corresponding to 9 primary constraints, we
require a weaker condition than the one in \citep{Dambrosio:2020wbi}.

\section{Perturbative analysis of the number of degrees of freedom \label{sec:pert}}

In this section, we consider the linear perturbations of the QSN action
(\ref{QSN_action}) in a cosmological background. This will provide
necessary conditions to count the number of degrees of freedom.

The spacetime metric in terms of the ADM variables is given by (\ref{eq:adm metric}).
The perturbations are taken around a spatially flat Friedmann-Robertson-Walker
background with metric $\mathrm{d}s^{2}=-\mathrm{d}t^{2}+a^{2}(t)\delta_{ij}\mathrm{d}x^{i}\mathrm{d}x^{j}$
and scalar field $\phi=\phi_{0}(t)$. The perturbed ADM variables
are parametrized by\footnote{Here we set $\bar{N}=1$ for simplicity. A subtle point is that since
time reparametrization has been fixed by the coincident gauge, $\bar{N}=1$
and a time-dependent $\bar{N}(t)$ correspond to physically inequivalent
background evolution. This is different from the case in generally
covariant theories, in which we can always set $\bar{N}=1$ using
time reparametrization symmetry.}
\begin{eqnarray}
N & = & e^{A},\label{ADM1}\\
N_{i} & = & aB_{i},\\
h_{ij} & = & a^{2}\delta_{ik}\left(e^{\bm{H}}\right)^{k}_{\phantom{k}j},\label{ADM3}
\end{eqnarray}
where $a=a\left(t\right)$ is the scale factor and the matrix exponential
is defined in the standard manner $\left(e^{\bm{H}}\right)^{i}_{\phantom{k}j}\equiv\delta^{i}_{\phantom{i}j}+H^{i}_{\phantom{i}j}+\frac{1}{2}H^{i}_{\phantom{i}k}H^{k}_{\phantom{i}j}+\cdots$
with $H^{i}_{\phantom{i}j}\equiv\delta^{ik}H_{kj}$. We further decompose
$B_{i}$ and $H_{ij}$ as
\begin{equation}
B_{i}\equiv\partial_{i}B+S_{i}
\end{equation}
and
\begin{equation}
H_{ij}\equiv2\zeta\delta_{ij}+\left(\partial_{i}\partial_{j}-\frac{1}{3}\delta_{ij}\partial^{2}\right)E+\partial_{i}F_{j}+\partial_{j}F_{i}+\gamma_{ij},
\end{equation}
with $\partial^{2}\equiv\delta^{ij}\partial_{i}\partial_{j}$, $S_{i}$
and $F_{i}$ satisfying $\delta^{ij}\partial_{i}S_{j}=\delta^{ij}\partial_{i}F_{j}=0$,
and $\gamma_{ij}$ satisfying $\delta^{ij}\gamma_{ij}=0$ and $\delta^{ij}\partial_{i}\gamma_{jk}=0$.
We thus have three types of perturbation modes:
\begin{eqnarray*}
A,B,\zeta,E & : & \text{scalar perturbations}\\
S_{i},F_{i} & : & \text{vector perturbations}\\
\gamma_{ij} & : & \text{tensor perturbations}
\end{eqnarray*}
The scalar field is perturbed as
\begin{equation}
\phi=\phi_{0}+\delta\phi,
\end{equation}
where $\phi_{0}(t)$ is the background value of the scalar field.
Working in the coincident gauge fixes the local diffeomorphism freedom
at the level of the action. Consequently, we cannot use the usual
gauge fixing to eliminate perturbation variables (up to the global
affine transformations that preserve $\Gamma^{\rho}{}_{\mu\nu}=0$).
We therefore keep all 10 metric perturbations together with the scalar
perturbation $\delta\phi$ in calculating the perturbations.

After straightforward yet tedious calculations, we can obtain the
action for the perturbation modes. The linear action for the perturbations
is
\begin{align}
S_{1}= & \int\mathrm{d}t\mathrm{d}^{3}x\Big[12a\dot{a}^{2}A\left(\bar{c}_{1}+3\bar{c}_{3}\right)-12a\dot{a}^{2}\delta\phi\left(\bar{c}_{1}'+3\bar{c}_{3}'\right)-36a\dot{a}^{2}\zeta\left(\bar{c}_{1}+3\bar{c}_{3}\right)\nonumber \\
 & -2a\dot{a}\Box B\left(2\bar{c}_{2}+3\bar{c}_{5}\right)-12a^{2}\dot{a}\dot{A}\left(2\bar{c}_{3}+\bar{c}_{5}\right)-24a^{2}\dot{a}\dot{\zeta}\left(\bar{c}_{1}+3\bar{c}_{3}\right)\nonumber \\
 & +\frac{1}{2}a^{3}\dot{\phi}^{2}_{0}\left(3\zeta-A\right)+a^{3}\delta\dot{\phi}\dot{\phi}_{0}+a^{3}V_{,\phi}\left(\phi_{0}\right)\delta\phi+a^{3}V\left(\phi_{0}\right)\left(A+3\zeta\right)\Big],
\end{align}
where $\bar{c}_{i}\coloneqq c_{i}\left(\phi_{0}\right)$ and $\bar{c}_{i}'\coloneqq\left.\frac{\partial c_{i}\left(\phi\right)}{\partial\phi}\right|_{\phi=\phi_{0}}$
with $i=1,2,3,4,5$. By making use of integrations by parts, $S_{1}=0$
yields the background equations of motion
\begin{equation}
12a\dot{a}^{2}\left(\bar{c}_{1}+7\bar{c}_{3}+2\bar{c}_{5}\right)+12a^{2}\dot{a}\left(2\dot{\bar{c}}_{3}+\dot{\bar{c}}_{5}\right)+12a^{2}\ddot{a}\left(2\bar{c}_{3}+\bar{c}_{5}\right)-a^{3}\left(\frac{1}{2}\dot{\phi}^{2}_{0}-V\left(\phi_{0}\right)\right)=0,\label{bgeom_1}
\end{equation}
\begin{equation}
4a\dot{a}^{2}\left(\bar{c}_{1}+3\bar{c}_{3}\right)+8a^{2}\dot{a}\left(\dot{\bar{c}}_{1}+3\dot{\bar{c}}_{3}\right)+8a^{2}\ddot{a}\left(\bar{c}_{1}+3\bar{c}_{3}\right)+a^{3}\left(\frac{1}{2}\dot{\phi}^{2}_{0}+V\left(\phi_{0}\right)\right)=0,
\end{equation}
and
\begin{equation}
-12a\dot{a}^{2}\left(\frac{\partial\bar{c}_{1}}{\partial\phi}+3\frac{\partial\bar{c}_{3}}{\partial\phi}\right)-3a^{2}\dot{a}\dot{\phi}_{0}-a^{3}\ddot{\phi}_{0}+a^{3}V_{,\phi}\left(\phi_{0}\right)=0.\label{bgeom_3}
\end{equation}
We can see that the background evolution gets modifications from the
coefficient $c_{i}$. However, such modifications may result in some
exotic evolution, as we will see below. Note that after imposing the
coincident gauge (which fixes the diffeomorphism freedom), the relation
among background equations that follows from manifest covariance is
no longer explicit in the gauge-fixed action. In a generally covariant
formulation, the equation of motion for the scalar field can be obtained
from the metric equations together with the contracted Bianchi identity
(equivalently, conservation of the energy-momentum tensor). In the
present gauge-fixed formulation, it is therefore convenient to treat
(\ref{bgeom_1})-(\ref{bgeom_3}) as independent equations.

We then consider the second order perturbations. For the sake of clarity,
we will separate the quadratic action into the scalar, vector, and
tensor sectors. After some manipulations, we find the quadratic action
for the scalar perturbations is given by
\begin{align}
S_{2,\mathrm{s}}= & \int\mathrm{d}t\mathrm{d}^{3}x\Big[-4a^{3}\dot{A}^{2}\left(\bar{c}_{1}+\bar{c}_{2}+\bar{c}_{3}+\bar{c}_{4}+\bar{c}_{5}\right)+\frac{1}{2}a^{3}\delta\dot{\phi}^{2}-12a^{3}\dot{\zeta}^{2}\left(\bar{c}_{1}+3\bar{c}_{3}\right)\nonumber \\
 & -12a^{3}\dot{A}\dot{\zeta}\left(2\bar{c}_{3}+\bar{c}_{5}\right)-a^{3}\dot{B}\Box\dot{B}\left(2\bar{c}_{1}+\bar{c}_{2}+\bar{c}_{4}\right)-\frac{2}{3}a^{3}\Box\dot{E}\Box\dot{E}\bar{c}_{1}+\cdots\Big],\label{S2_sca}
\end{align}
where ``$\cdots$'' denotes terms that contain at most a single
time derivative and thus are irrelevant to counting the number of
degrees of freedom. The full expression of $S_{2,\mathrm{s}}$ can
be found in Appendix \ref{app:S2conc}. The quadratic action for the
vector perturbations is given by
\begin{align}
S_{2,\mathrm{v}}= & \int\mathrm{d}t\mathrm{d}^{3}x\Big[a\dot{a}^{2}S_{i}S^{i}\left(2\bar{c}_{1}+\bar{c}_{2}+\bar{c}_{4}\right)+2a\bar{c}_{1}S_{i}\Box S^{i}+a^{2}\bar{c}_{2}S^{i}\Box\dot{F}_{i}+\frac{1}{4}a\Box F^{i}\Box F_{i}\left(2\bar{c}_{1}+\bar{c}_{2}+\bar{c}_{4}\right)\nonumber \\
 & +a\dot{a}\bar{c}_{4}S^{i}\Box F_{i}-a^{2}\bar{c}_{4}\dot{S}_{i}\Box F^{i}-2a^{2}\dot{a}\left(2\bar{c}_{1}+\bar{c}_{2}+\bar{c}_{4}\right)S^{i}\dot{S}_{i}+a^{3}\left(2\bar{c}_{1}+\bar{c}_{2}+\bar{c}_{4}\right)\dot{S}_{i}\dot{S}^{i}+\frac{1}{2}a^{3}\bar{c}_{1}\dot{F}_{i}\Box\dot{F}^{i}\Big],\label{S2_vec}
\end{align}
and the quadratic action for the tensor perturbations is given by
\begin{equation}
S_{2,\mathrm{t}}=\int\mathrm{d}t\mathrm{d}^{3}x\left(-a^{3}\bar{c}_{1}\dot{\gamma}_{ij}\dot{\gamma}^{ij}-a\bar{c}_{1}\gamma^{ij}\Box\gamma_{ij}\right).\label{S2_ten}
\end{equation}
From (\ref{S2_sca})-(\ref{S2_ten}), in general all the perturbation
modes acquire a kinetic term and thus become dynamical if no specific
conditions are assumed. 

In the rest of this section, we will discuss the various cases that
we have classified in Sec. \ref{sec:pricons}. In particular, we will
show the part in the quadratic action that is (or will be after solving
the constraints) quadratic in velocity terms, which is relevant to
counting the number of dynamical degrees of freedom. 

\subsection{Cases III, IV, and VII}

In all three cases, the background equations are exactly the same,
which are given by
\begin{align}
-\frac{1}{2}a^{3}\dot{\phi}^{2}_{0}+a^{3}\bar{V} & =0,\label{bgeom_up_1}\\
\frac{3}{2}a^{3}\dot{\phi}^{2}_{0}+3a^{3}\bar{V} & =0,\label{bgeom_up_2}\\
a^{3}\bar{V}_{,\phi}-3a^{2}\dot{a}\dot{\phi}_{0}-a^{3}\ddot{\phi}_{0} & =0.\label{bgeom_up_3}
\end{align}
Solving (\ref{bgeom_up_1})-(\ref{bgeom_up_3}) enforces $\bar{V}=0$
and $\dot{\phi}_{0}=0$. In particular, the background equations in
these cases do not provide a standard Friedmann evolution for the
scale factor $a(t)$. This reflects the degeneracy that the trace
part of $\dot{h}_{ij}$ (equivalently $K_{ij}$) cannot be solved
for, as can be seen in Appendix \ref{app:pri_cons}. As a result,
a perturbative analysis around an expanding FLRW background is not
well-defined in these cases, and we will not consider them further.

\subsection{Case II \label{subsec:pert_case2}}

This case corresponds to $c_{5}=-2c_{3}$ and $c_{4}=c_{3}-c_{1}-c_{2}$.
In this case, the quadratic action for the scalar perturbations becomes
\begin{align}
S_{2,\mathrm{s}}\supset & \int\mathrm{d}t\mathrm{d}^{3}x\bigg\{\frac{1}{2}a^{3}\delta\dot{\phi}^{2}-12a^{3}\left(\bar{c}_{1}+3\bar{c}_{3}\right)\dot{\zeta}^{2}-a^{3}\left(\bar{c}_{1}+\bar{c}_{3}\right)\dot{B}\Box\dot{B}-\frac{2}{3}a^{3}\bar{c}_{1}\Box\dot{E}\Box\dot{E}\nonumber \\
 & -4aA\Box A\left(\bar{c}_{1}+\bar{c}_{3}\right)+24a^{2}\dot{a}\dot{\zeta}A\left(\bar{c}_{1}+3\bar{c}_{3}\right)-4a^{2}A\Box\dot{B}\left(\bar{c}_{1}+\bar{c}_{3}\right)\nonumber \\
 & -a^{3}\dot{\phi}_{0}A\delta\dot{\phi}+\left[-12a\dot{a}^{2}\left(\bar{c}_{1}+3\bar{c}_{3}\right)+\frac{1}{2}a^{3}\dot{\phi}^{2}_{0}\right]A^{2}\bigg\},\label{S2_sca_case2}
\end{align}
where $\square\equiv\partial_{i}\partial^{i}$. Here and in the following,
``$\supset$'' indicates that we have neglected terms irrelevant
to the kinetic terms (after solving the auxiliary variables). In deriving
(\ref{S2_sca_case2}) we have used the background equations of motion
to simplify the quadratic action. In this case, the kinetic term of
$A$ vanishes, which implies that $A$ becomes an auxiliary variable.
Moreover, one can check that the quadratic actions for the vector
and tensor perturbations receive no change and thus are the same as
in (\ref{S2_vec}) and (\ref{S2_ten}).

To solve the auxiliary variable $A$, it is convenient to rewrite
the quadratic action (\ref{S2_sca_case2}) in Fourier space\footnote{Here and below, the perturbation variables denote their Fourier transformation.
In particular, terms such as $\delta\dot{\phi}^{2}$, $\dot{\zeta}^{2}$
are shorthand for $\delta\dot{\phi}_{\bm{k}}\delta\dot{\phi}_{-\bm{k}}$,
$\dot{\zeta}_{\bm{k}}\dot{\zeta}_{-\bm{k}}$ etc.}, 
\begin{align}
S_{2,\mathrm{s}}\supset & \int\mathrm{d}t\frac{\mathrm{d}^{3}k}{(2\pi)^{3}}\bigg\{\frac{1}{2}a^{3}\delta\dot{\phi}^{2}-12a^{3}\left(\bar{c}_{1}+3\bar{c}_{3}\right)\dot{\zeta}^{2}+a^{3}\left(\bar{c}_{1}+\bar{c}_{3}\right)\bm{k}^{2}\dot{B}^{2}-\frac{2}{3}a^{3}\bar{c}_{1}\bm{k}^{4}\dot{E}^{2}\nonumber \\
 & +4a\bm{k}^{2}A^{2}\left(\bar{c}_{1}+\bar{c}_{3}\right)+24a^{2}\dot{a}\dot{\zeta}A\left(\bar{c}_{1}+3\bar{c}_{3}\right)+4a^{2}\bm{k}^{2}A\dot{B}\left(\bar{c}_{1}+\bar{c}_{3}\right)\nonumber \\
 & -a^{3}\dot{\phi}_{0}A\delta\dot{\phi}+\left[-12a\dot{a}^{2}\left(\bar{c}_{1}+3\bar{c}_{3}\right)+\frac{1}{2}a^{3}\dot{\phi}^{2}_{0}\right]A^{2}\bigg\}.\label{S2_case2_mom}
\end{align}
Then we can obtain the constraint equation of $A$ with the solution
\begin{align}
A\supset & -\frac{4a^{2}\bm{k}^{2}\left(\bar{c}_{1}+\bar{c}_{3}\right)}{g}\dot{B}-\frac{24a^{2}\dot{a}\left(\bar{c}_{1}+3\bar{c}_{3}\right)}{g}\dot{\zeta}+\frac{a^{3}\dot{\phi}_{0}}{g}\dot{\delta\phi},\label{solA_case2}
\end{align}
with 
\begin{equation}
g=-24a\dot{a}\left(\bar{c}_{1}+3\bar{c}_{3}\right)+a^{3}\dot{\phi}_{0}+8a\bm{k}^{2}\left(\bar{c}_{1}+\bar{c}_{3}\right),\label{g_def}
\end{equation}
which depends on the scale factor and background values of coefficients
$c_{1}$ and $c_{3}$. Plugging the solution (\ref{solA_case2}) into
(\ref{S2_case2_mom}), we can obtain the quadratic action for the
scalar perturbations $\zeta$, $E$, $B$, and $\delta\phi$. 

For our purpose, we focus on the kinetic terms. The Hessian matrix
of perturbation variables $\left\{ \zeta,B,E,\delta\phi\right\} $
is given by
\begin{equation}
\bm{H}=\left(\begin{array}{cccc}
H_{\zeta\zeta} & H_{\zeta B} & 0 & H_{\zeta\delta\phi}\\
H_{\zeta B} & H_{BB} & 0 & H_{B\delta\phi}\\
0 & 0 & H_{EE} & 0\\
H_{\zeta\delta\phi} & H_{B\delta\phi} & 0 & H_{\delta\phi\delta\phi}
\end{array}\right),\label{Hessian_case2}
\end{equation}
 where the nonvanishing entries are
\begin{equation}
H_{\zeta\zeta}=-12a^{3}\bigg\{\left(\bar{c}_{1}+3\bar{c}_{3}\right)-\frac{24}{g}\left[2a\dot{a}^{2}\left(\bar{c}^{2}_{1}+4\bar{c}_{1}\bar{c}_{3}+3\bar{c}^{2}_{3}\right)+a\dot{a}^{2}\left(\bar{c}_{1}+\bar{c}_{3}\right)^{2}\right]\bigg\},
\end{equation}
\begin{equation}
H_{\zeta B}=-\frac{48}{g}\bm{k}^{2}a^{4}\dot{a}\left(\bar{c}_{1}+\bar{c}_{3}\right)^{2},
\end{equation}
\begin{equation}
H_{\zeta\delta\phi}=\frac{12}{g}a^{5}\dot{a}\left(\bar{c}_{1}+\bar{c}_{3}\right)\dot{\phi}_{0},
\end{equation}
\begin{equation}
H_{BB}=-\bm{k}^{2}a^{3}\left(\bar{c}_{1}+\bar{c}_{3}\right)-\frac{8}{g}\bm{k}^{4}a^{4}\left(\bar{c}_{1}+\bar{c}_{3}\right)^{2},
\end{equation}
\begin{equation}
H_{B\delta\phi}=\frac{2}{g}\bm{k}^{2}a^{5}\left(\bar{c}_{1}+\bar{c}_{3}\right)\dot{\phi}_{0},
\end{equation}
\begin{equation}
H_{EE}=-\frac{2}{3}\bm{k}^{4}a^{2}\bar{c}_{1},
\end{equation}
and
\begin{equation}
H_{\delta\phi\delta\phi}=\frac{a^{3}}{2}-\frac{a^{6}\dot{\phi}^{2}_{0}}{2g}.
\end{equation}
One can check that the Hessian is not degenerate, $\det\bm{H}\neq0$,
and thus all the variables are dynamical. 

In addition, the vector modes $S^{i}$, $F^{i}$, and the tensor modes
$\gamma^{ij}$ are all dynamical, since all of which acquire kinetic
terms. Therefore, the number of degrees of freedom is at least 10
(4 scalar modes, 4 vector modes, and 2 tensor modes) in this case.
In the next section, by evaluating the Poisson brackets of primary
constraints, we will see that there are two constraints in the phase
space, which is consistent with the perturbation analysis.

\subsection{Case V \label{subsec:pert_case5}}

This case corresponds to $c_{4}=-2c_{1}-c_{2}$. In this case, $B$
and $S^{i}$ have no kinetic terms and become auxiliary variables,
which can be solved in terms of other dynamical variables. We first
consider the scalar sector. The quadratic action for the scalar perturbations
is 
\begin{align}
S_{2,\mathrm{s}}\supset & \int\mathrm{d}t\frac{\mathrm{d}^{3}k}{\left(2\pi\right)^{3}}\bigg[4a^{3}\left(\bar{c}_{1}-\bar{c}_{3}-\bar{c}_{5}\right)\dot{A}^{2}+\frac{1}{2}a^{3}\delta\dot{\phi}^{2}-12a^{3}\left(\bar{c}_{1}+3\bar{c}_{3}\right)\dot{\zeta}^{2}-\frac{2}{3}a^{3}\bar{c}_{1}\boldsymbol{k}^{4}\dot{E}^{2}\nonumber \\
 & -12a^{3}\left(2\bar{c}_{3}+\bar{c}_{5}\right)\dot{A}\dot{\zeta}-4a^{2}\left(2\bar{c}_{1}-\bar{c}_{5}\right)\boldsymbol{k}^{2}\dot{A}B-4a^{2}\left(2\bar{c}_{1}-3\bar{c}_{3}\right)\boldsymbol{k}^{2}\dot{\zeta}B+\frac{8}{3}a^{2}\bar{c}_{1}\boldsymbol{k}^{4}\dot{E}B\biggr].
\end{align}
By varying it with respect to $B$, we can obtain the constraint equation
\begin{equation}
\left(2\bar{c}_{1}-\bar{c}_{5}\right)\dot{A}+\left(2\bar{c}_{1}-3\bar{c}_{5}\right)\dot{\zeta}-\frac{2}{3}\bar{c}_{1}\boldsymbol{k}^{2}\dot{E}+\cdots=0,\label{conseq_case5}
\end{equation}
where ``$\cdots$'' denotes terms involving no temporal derivative.
Equation (\ref{conseq_case5}) is a nonholonomic constraint among $A$, $\zeta$,
and $E$, which by itself does not eliminate any degree of freedom
\citep{Gao:2019lpz}. The Hessian matrix of perturbation variables
$\left\{ \zeta,A,E,\delta\phi\right\} $ is given by
\begin{equation}
\bm{H}=\left(\begin{array}{cccc}
H_{\zeta\zeta} & H_{\zeta A} & 0 & 0\\
H_{\zeta A} & H_{AA} & 0 & 0\\
0 & 0 & H_{EE} & 0\\
0 & 0 & 0 & H_{\delta\phi\delta\phi}
\end{array}\right),\label{Hessian_case5}
\end{equation}
where the nonvanishing entries are
\begin{equation}
H_{\zeta\zeta}=-12a^{3}\left(\bar{c}_{1}+3\bar{c}_{3}\right),
\end{equation}
\begin{equation}
H_{\zeta A}=-6a^{3}\left(2\bar{c}_{3}+\bar{c}_{5}\right),
\end{equation}
\begin{equation}
H_{AA}=4a^{3}\left(\bar{c}_{1}-\bar{c}_{3}-\bar{c}_{5}\right),
\end{equation}
\begin{equation}
H_{EE}=-\frac{2}{3}a^{3}\bar{c}_{1}\boldsymbol{k}^{4},
\end{equation}
and
\begin{equation}
H_{\delta\phi\delta\phi}=\frac{1}{2}a^{3}.
\end{equation}
Since the Hessian matrix is not degenerate, there are 4 DOFs in the
scalar sector.

We now turn to the vector sector. The quadratic action of vector perturbations
is given by 
\begin{align}
S_{2,\mathrm{v}}= & \int\mathrm{d}t\mathrm{d}^{3}x\Big[2a\bar{c}_{1}S_{i}\Box S^{i}+a^{2}\bar{c}_{2}S^{i}\Box\dot{F}_{i}\nonumber \\
 & -a\dot{a}\left(2\bar{c}_{1}+\bar{c}_{2}\right)S^{i}\Box F_{i}+a^{2}\left(2\bar{c}_{1}+\bar{c}_{2}\right)\dot{S}_{i}\Box F^{i}+\frac{1}{2}a^{3}\bar{c}_{1}\dot{F}_{i}\Box\dot{F}^{i}\Big],\label{S2_vec_case5}
\end{align}
from which we can solve the auxiliary variable $S^{i}$ in terms of
$F^{i}$ and get 
\begin{equation}
S_{i}=\frac{a\dot{F}_{i}}{2}.
\end{equation}
Interestingly, after plugging the solution into  (\ref{S2_vec_case5}),
the kinetic term of $F^{i}$ vanishes identically. As a result, in
this case there are no vector DOFs at the linear order in perturbations.
The quadratic action for the tensor modes (\ref{S2_ten}) is not altered.
We can conclude that in this case there are 6 physical DOFs at the
linear order in perturbations.

\subsection{Case VI \label{subsec:pert_case6}}

This case corresponds to $c_{3}=-c_{1}$, $c_{4}=-2c_{1}-c_{2}$, and
$c_{5}=2c_{1}$. As in the previous case, $A$, $B$, and $S^{i}$
have no kinetic terms and become auxiliary variables. The quadratic
action for the scalar perturbations becomes
\begin{align}
S_{2,s}\supset & \int\mathrm{d}t\frac{\mathrm{d}^{3}k}{\left(2\pi\right)^{3}}\biggl[\left(24a\dot{a}^{2}\bar{c}_{1}+\frac{1}{2}a^{3}\dot{\phi}^{2}_{0}\right)A^{2}+\frac{1}{2}a^{3}\delta\dot{\phi}^{2}+24a^{3}\bar{c}_{1}\dot{\zeta}^{2}\nonumber \\
 & +8a\dot{a}\bar{c}_{2}\boldsymbol{k}^{2}AB+4a^{2}\left(\dot{c}_{1}+\dot{c}_{2}\right)\boldsymbol{k}^{2}AB+16a^{2}\bar{c}_{1}\boldsymbol{k}^{2}\dot{\zeta}B\nonumber \\
 & -48a^{2}\dot{a}\bar{c}_{1}A\dot{\zeta}-a^{3}\dot{\phi}_{0}A\delta\dot{\phi}+\frac{8}{3}a^{2}\bar{c}_{1}\boldsymbol{k}^{4}\dot{E}B-\frac{2}{3}a^{3}\bar{c}_{1}\boldsymbol{k}^{4}\dot{E}^{2}\biggr].\label{S2_sca_case6}
\end{align}
We can solve $A$ and $B$ in terms of other variables as
\begin{equation}
A\supset-\frac{16a^{2}\bar{c}_{1}}{g}\dot{\zeta}-\frac{8a^{2}\bar{c}_{1}\boldsymbol{k}^{2}}{3g}\dot{E}
\end{equation}
and
\begin{align}
B\supset & \frac{16\bar{c}_{1}a^{2}(48\bar{c}_{1}a\dot{a}^{2}+a^{3}\dot{\phi}_{0}{}^{2}+3\dot{a}g)}{\boldsymbol{k}^{2}g^{2}}\dot{\zeta}\nonumber \\
 & +\frac{8\bar{c}_{1}a^{3}(48\bar{c}_{1}\dot{a}^{2}+a^{2}\dot{\phi}_{0}{}^{2})}{3g^{2}}\dot{E}+\frac{a^{3}\dot{\phi}_{0}}{\boldsymbol{k}^{2}g}\delta\dot{\phi},
\end{align}
where $g\coloneqq8a\dot{a}\bar{c}_{2}+4a^{2}\left(\bar{c}_{1}{}^{\prime}+\bar{c}_{2}{}^{\prime}\right)\dot{\phi}_{0}$.
Plugging these solutions into the quadratic action (\ref{S2_sca_case6}),
the Hessian matrix of the perturbation variables $\left\{ \zeta,E,\delta\phi\right\} $
is
\begin{equation}
\bm{H}=\left(\begin{array}{ccc}
H_{\zeta\zeta} & H_{\zeta E} & H_{\zeta\delta\phi}\\
H_{\zeta E} & H_{EE} & H_{E\delta\phi}\\
H_{\zeta\delta\phi} & H_{E\delta\phi} & H_{\delta\phi\delta\phi}
\end{array}\right),\label{Hessian_case6}
\end{equation}
where
\begin{align}
H_{\zeta\zeta}= & -\frac{8\bar{c_{1}}a^{3}}{g^{3}}\Biggl[768\bar{c}_{1}{}^{2}a^{2}\dot{a}^{2}\left(16\bar{c}_{2}a\dot{a}+8a^{2}(\bar{c}_{1}{}^{\prime}+\bar{c}_{2}{}^{\prime})\dot{\phi}_{0}-3g\right)-3g^{3}\nonumber \\
 & +16\bar{c}_{1}a\left(8a^{5}(\bar{c}_{1}{}^{\prime}+\bar{c}_{2}{}^{\prime})\dot{\phi}_{0}{}^{3}+24a^{2}(\bar{c}_{1}{}^{\prime}+\bar{c}_{2}{}^{\prime})\dot{a}\dot{\phi}_{0}g\right)\nonumber \\
 & +16\bar{c}_{1}a\left(-3a^{3}\dot{\phi}_{0}{}^{2}g-12\dot{a}g^{2}+16\bar{c}_{2}a\dot{a}(a^{3}\dot{\phi}_{0}{}^{2}+3\dot{a}g)\right)\Biggr],
\end{align}
\begin{align}
H_{\zeta E}= & -\frac{64\bar{c}_{1}{}^{2}a^{4}\boldsymbol{k}^{2}}{3g^{3}}\Biggl[8a^{5}c_{1}{}^{\prime}\dot{\phi}_{0}{}^{3}+8a^{5}c_{2}{}^{\prime}\dot{\phi}_{0}{}^{3}+48\bar{c}_{1}a\dot{a}^{2}\left(16\bar{c}_{2}a\dot{a}+8a^{2}(\bar{c}_{1}{}^{\prime}+\bar{c}_{2}{}^{\prime})\dot{\phi}_{0}-3g\right)\nonumber \\
 & +12a^{2}c_{1}{}^{\prime}\dot{a}\dot{\phi}_{0}g+12a^{2}c_{2}{}^{\prime}\dot{a}\dot{\phi}_{0}g-3a^{3}\dot{\phi}_{0}{}^{2}g-6\dot{a}g^{2}+8\bar{c}_{2}a\dot{a}\left(2a^{3}\dot{\phi}_{0}{}^{2}+3\dot{a}g\right)\Biggr],
\end{align}
\begin{equation}
H_{\zeta\delta\phi}=-\frac{16}{g^{2}}\bar{c}_{1}a^{5}\dot{\phi}_{0}(4\bar{c}_{2}a\dot{a}+2a^{2}(\bar{c}_{1}{}^{\prime}+\bar{c}_{2}{}^{\prime})\dot{\phi}_{0}-g),
\end{equation}
\begin{align}
H_{EE}= & -\frac{2\bar{c}_{1}a^{3}\boldsymbol{k}^{4}}{9g^{3}}\Biggl[768\bar{c}_{1}{}^{2}a^{2}\dot{a}^{2}\left(16\bar{c}_{2}a\dot{a}+8a^{2}(\bar{c}_{1}{}^{\prime}+\bar{c}_{2}{}^{\prime})\dot{\phi}_{0}-3g\right)\nonumber \\
 & +16\bar{c}_{1}a^{4}\dot{\phi}_{0}{}^{2}\left(16\bar{c}_{2}a\dot{a}+8a^{2}(\bar{c}_{1}{}^{\prime}+\bar{c}_{2}{}^{\prime})\dot{\phi}_{0}-3g\right)+3g^{3}\Biggr],
\end{align}
\begin{equation}
H_{E\delta\phi}=-\frac{8\bar{c}_{1}\boldsymbol{k}^{2}}{3g^{2}}a^{5}\dot{\phi}_{0}(4\bar{c}_{2}a\dot{a}+2a^{2}(\bar{c}_{1}{}^{\prime}+\bar{c}_{2}{}^{\prime})\dot{\phi}_{0}-g),
\end{equation}
and
\begin{equation}
H_{\delta\phi\delta\phi}=\frac{1}{2}a^{3}.
\end{equation}
Again, since generally the Hessian matrix is not degenerate, all the
variables $\zeta$, $E$, and $\delta\phi$ are dynamical. 

Note since STEGR is a special case of Case VI, we can calculate the
perturbations of STEGR as a consistency check. Taking $c_{1}=1/4$,
$c_{2}=-1/2$, $c_{3}=-1/4$, $c_{4}=0$, and $c_{5}=1/2$, the Hessian
matrix (\ref{subsec:pert_case6}) reduces to
\begin{equation}
\boldsymbol{H}_{\mathrm{STEGR}}=\frac{1}{2}a^{3}\left(\begin{array}{ccc}
\frac{a^{2}\dot{\phi}_{0}{}^{2}}{\dot{a}^{2}} & \frac{a^{2}\boldsymbol{k}^{2}\dot{\phi}_{0}{}^{2}}{6\dot{a}^{2}} & -\frac{a\dot{\phi}_{0}}{\dot{a}}\\
\frac{a^{2}\boldsymbol{k}^{2}\dot{\phi}_{0}{}^{2}}{6\dot{a}^{2}} & \frac{a^{2}\boldsymbol{k}^{4}\dot{\phi}_{0}{}^{2}}{36\dot{a}^{2}} & -\frac{a\boldsymbol{k}^{2}\dot{\phi}_{0}}{6\dot{a}}\\
-\frac{a\dot{\phi}_{0}}{\dot{a}} & -\frac{a\boldsymbol{k}^{2}\dot{\phi}_{0}}{6\dot{a}} & 1
\end{array}\right).
\end{equation}
It is clear that the Hessian matrix becomes degenerate in STEGR with
$\mathrm{rank}\bm{H}_{\mathrm{STEGR}}=1$. This means there is only
one dynamical degree of freedom for the scalar perturbations as expected.

Then we turn to the vector case. As in case V, we find that the kinetic
terms of the vector perturbations identically vanish after solving
the variable $S^{i}$. To conclude, in case VI, generally there are
5 DOFs (3 scalar and 2 tensor) at the linear order in perturbations.
In the special case of STEGR, there are 3 DOFs (1 scalar and 2 tensor)
at the linear order in perturbations. As we will see in Sec. \ref{sec:sec_cons},
there is a discrepancy between the nonperturbative Hamiltonian analysis
and the number of modes visible at quadratic order on FLRW. This suggests
that some modes have vanishing quadratic kinetic terms and thus are
strongly coupled on FLRW background. Generally, such modes will acquire
kinetic terms beyond quadratic order on FLRW and/or become dynamical
already at linear order on less symmetric backgrounds.

\section{Secondary constraints and degrees of freedom \label{sec:sec_cons}}

In the previous section, we have made a perturbative analysis of the
number of DOFs of the QSN theory, with different choices of coefficients.
The numbers of DOFs determined with this method, however, are merely
the lower bounds of the numbers of DOFs, since perturbation modes
may manifest themselves only on nonlinear order in perturbations.
Therefore, a nonperturbative analysis of the number of DOFs based
on the Hamiltonian constraint analysis is necessary. This is the task
of this section.

Our starting point is the ADM-decomposed action (\ref{action_ADM}).
It is convenient to expand the extrinsic curvature in terms of $\dot{h}_{ij}$
explicitly and write 
\begin{align}
S= & \int\mathrm{d}t\mathrm{d}^{3}x\sqrt{h}\bigg\{ N\overset{3}{\mathcal{Q}}-\frac{c_{1}h^{ij}h^{kl}\dot{h}_{ik}\dot{h}_{jl}}{N}-\frac{c_{3}h^{ij}h^{kl}\dot{h}_{ij}\dot{h}_{kl}}{N}-\frac{2(2c_{3}+c_{5})h^{ij}\dot{h}_{ij}\dot{N}}{N^{2}}\nonumber \\
 & -\frac{4(c_{1}+c_{2}+c_{3}+c_{4}+c_{5})\dot{N}^{2}}{N^{3}}+\frac{(2c_{1}+c_{2}+c_{4})h_{ij}\dot{N}^{i}\dot{N}^{j}}{N^{3}}\nonumber \\
 & +\left[\frac{2c_{1}h^{jl}h^{km}N^{i}\partial_{i}h_{lm}}{N}+\frac{2c_{3}h^{jk}h^{lm}N^{i}\partial_{i}h_{lm}}{N}+\frac{2(2c_{3}+c_{5})h^{jk}N^{i}\partial_{i}N}{N^{2}}-\frac{2c_{2}h^{ik}\partial_{i}N^{j}}{N}-\frac{c_{5}h^{jk}\partial_{i}N^{i}}{N}\right]\dot{h}_{jk}\nonumber \\
 & +\left[\frac{8(c_{1}+c_{2}+c_{3}+c_{4}+c_{5})N^{i}\partial_{i}N}{N^{3}}-\frac{2(2c_{4}+c_{5})\partial_{i}N^{i}}{N^{2}}+\frac{2(2c_{3}+c_{5})h^{jk}N^{i}\partial_{i}h_{jk}}{N^{2}}\right]\dot{N}\nonumber \\
 & +\left[-\frac{c_{5}h^{jk}\partial_{i}h_{jk}}{N}-\frac{2(2c_{2}+c_{5})\partial_{i}N}{N^{2}}-\frac{2c_{4}h^{jk}\partial_{k}h_{ij}}{N}-\frac{2(2c_{1}+c_{2}+c_{4})h_{ik}N^{j}\partial_{j}N^{k}}{N^{3}}\right]\dot{N}^{i}\nonumber \\
 & +4c_{3}h^{ij}h^{kl}\partial_{i}N\partial_{j}h_{kl}+\frac{2(2c_{2}+c_{5})N^{i}\partial_{i}N^{j}\partial_{j}N}{N^{2}}-\frac{c_{2}\partial_{i}N^{j}\partial_{j}N^{i}}{N}\nonumber \\
 & -\frac{4(c_{1}+c_{2}+c_{3}+c_{4}+c_{5})(N^{i})^{2}(\partial_{i}N)^{2}}{N^{3}}-\frac{2(2c_{3}+c_{5})h^{kl}N^{i}N^{j}\partial_{i}N\partial_{j}h_{kl}}{N^{2}}\nonumber \\
 & +\frac{c_{5}h^{kl}N^{i}\partial_{i}N^{j}\partial_{j}h_{kl}}{N}-\frac{c_{1}h^{kl}h^{mn}N^{i}N^{j}\partial_{i}h_{km}\partial_{j}h_{ln}}{N}-\frac{c_{3}h^{kl}h^{mn}N^{i}N^{j}\partial_{i}h_{kl}\partial_{j}h_{mn}}{N}\nonumber \\
 & +\frac{4(c_{1}+c_{3})h^{ij}\partial_{i}N\partial_{j}N}{N}+\frac{c_{5}h^{kl}N^{i}\partial_{i}h_{kl}\partial_{j}N^{j}}{N}+\frac{2(2c_{4}+c_{5})N^{i}\partial_{i}N\partial_{j}N^{j}}{N^{2}}\nonumber \\
 & -\frac{c_{4}\partial_{i}N^{i}\partial_{j}N^{j}}{N}+\frac{(2c_{1}+c_{2}+c_{4})h_{kl}N^{i}N^{j}\partial_{i}N^{k}\partial_{j}N^{l}}{N^{3}}+\frac{2c_{2}h^{kl}N^{i}\partial_{i}h_{jl}\partial_{k}N^{j}}{N}\nonumber \\
 & +2c_{5}h^{ij}h^{kl}\partial_{i}N\partial_{l}h_{jk}+\frac{2c_{4}h^{kl}N^{i}\partial_{i}N^{j}\partial_{l}h_{jk}}{N}-\frac{2c_{1}h_{ik}h^{jl}\partial_{j}N^{i}\partial_{l}N^{k}}{N}\nonumber \\
 & +\frac{\dot{\phi}^{2}}{2N}-\frac{N^{i}\partial_{i}\phi\dot{\phi}}{N}-\frac{1}{2}Nh^{ij}\partial_{i}\phi\partial_{j}\phi+\frac{1}{2}\frac{N^{i}N^{j}\partial_{i}\phi\partial_{j}\phi}{N}+NV\left(\phi\right)\bigg\}.
\end{align}
In the following, we will perform the Hamiltonian constraint analysis
of various cases discussed in the previous section. Since cases III,
IV, and VII acquire no consistent cosmological background solutions,
in the following we will focus on cases II, V, and VI.

\subsection{Case II}

First, we consider case II with $c_{5}=-2c_{3}$ and $c_{4}=c_{3}-c_{1}-c_{2}$.
With this choice of coefficients, the conjugate momenta are 
\begin{align}
\pi^{ij}= & \sqrt{h}\bigg(-\frac{2c_{3}\dot{h}_{kl}h^{kl}h^{ij}}{N}-\frac{2c_{1}\dot{h}_{kl}h^{ik}h^{jl}}{N}+\frac{2c_{3}h^{kl}h^{ij}N^{m}\partial_{m}h_{kl}}{N}\nonumber \\
 & +\frac{2c_{1}h^{ik}h^{jl}N^{m}\partial_{m}h_{kl}}{N}+\frac{2c_{3}h^{ij}\partial_{k}N^{k}}{N}-\frac{c_{2}h^{jk}\partial_{k}N^{i}}{N}-\frac{c_{2}h^{ik}\partial_{k}N^{j}}{N}\bigg),\label{case2_piij2}
\end{align}
\begin{align}
\pi_{i}= & \sqrt{h}\bigg[\frac{2\left(c_{1}+c_{3}\right)\dot{N}^{j}h_{ij}}{N^{3}}-\frac{2\left(c_{1}+c_{3}\right)h_{ij}N^{k}\partial_{k}N^{j}}{N^{3}}+\frac{2\left(c_{1}+c_{2}-c_{3}\right)h^{jk}\partial_{k}h_{ij}}{N}\nonumber \\
 & +\frac{2c_{3}h^{jk}\partial_{i}h_{jk}}{N}+\frac{\left(-4c_{2}+4c_{3}\right)\partial_{i}N}{N^{2}}\bigg],\label{case2_pi2}
\end{align}
\begin{equation}
\pi_{N}=\frac{4\left(c_{1}+c_{2}\right)\sqrt{h}}{N^{2}}\partial_{i}N^{i},\label{case2_pin2}
\end{equation}
and
\begin{equation}
\pi_{\phi}=\sqrt{h}\left(\frac{\dot{\phi}}{N}-\frac{N^{i}\partial_{i}\phi}{N}\right).\label{case2_piphi}
\end{equation}
It is clear that the only primary constraint is 
\begin{equation}
\phi^{(\mathrm{II})}_{N}\coloneqq\pi_{N}-4\sqrt{h}\left(c_{1}+c_{2}\right)\frac{\partial_{i}N^{i}}{N^{2}}.
\end{equation}

According to (\ref{case2_piij2})-(\ref{case2_piphi}), we can solve
$\dot{h}_{ij}$, $\dot{N}^{i}$, and $\dot{\phi}$ as 
\begin{align}
\dot{h}_{ij}= & -\frac{N\pi_{ij}}{2c_{1}\sqrt{h}}+\frac{c_{3}h_{ij}N\pi}{2c_{1}\left(c_{1}+3c_{3}\right)\sqrt{h}}+N^{k}\partial_{k}h_{ij}-\frac{c_{2}h_{kj}\partial_{i}N^{k}}{2c_{1}}\nonumber \\
 & -\frac{c_{2}h_{ki}\partial_{j}N^{k}}{2c_{1}}+\frac{c_{3}h_{ij}\partial_{k}N^{k}}{c_{1}}-\frac{c_{3}\left(-c_{2}+3c_{3}\right)h_{ij}\partial_{k}N^{k}}{c_{1}\left(c_{1}+3c_{3}\right)},\label{eq:case2 solved doth}
\end{align}
\begin{align}
\dot{N}^{i}= & \frac{N^{3}\pi^{i}}{2\left(c_{1}+c_{3}\right)\sqrt{h}}-\frac{2\left(-c_{2}+c_{3}\right)h^{ij}N\partial_{j}N}{c_{1}+c_{3}}+N^{j}\partial_{j}N^{i}\nonumber \\
 & -\frac{\left(c_{1}+c_{2}-c_{3}\right)h^{ij}h^{kl}N^{2}\partial_{k}h_{jl}}{c_{1}+c_{3}}-\frac{c_{3}h^{il}h^{jk}N^{2}\partial_{l}h_{jk}}{c_{1}+c_{3}},\label{eq:case2 solved dotshift}
\end{align}
and
\begin{equation}
\dot{\phi}=\frac{N\pi_{\phi}}{\sqrt{h}}+N^{i}\partial_{i}\phi.\label{eq:case2 solved phidot}
\end{equation}
By making use of all these equations above, we can obtain the canonical
Hamiltonian density 
\begin{align}
\mathcal{H}^{(\mathrm{II})}_{0} & =\frac{c_{3}N\pi^{2}}{(4c^{2}_{1}+12c_{1}c_{3})\sqrt{h}}-\frac{N\pi_{ij}\pi^{ij}}{4c_{1}\sqrt{h}}+\frac{N^{3}\pi_{i}\pi^{i}}{(4c_{1}+4c_{3})\sqrt{h}}+\frac{N\pi^{2}_{\phi}}{2\sqrt{h}}\nonumber \\
 & +N^{i}\pi^{jk}\partial_{i}h_{jk}+\frac{(c_{1}+c_{2})c_{3}\pi\partial_{i}N^{i}}{c_{1}(c_{1}+3c_{3})}-\frac{c_{2}h_{ik}\pi^{jk}\partial_{j}N^{i}}{c_{1}}\nonumber \\
 & +N^{i}\pi_{j}\partial_{i}N^{j}-\frac{c_{3}h^{ij}h^{kl}N^{2}\pi_{i}\partial_{j}h_{kl}}{c_{1}+c_{3}}+\frac{2(c_{2}-c_{3})h^{ij}N\pi_{i}\partial_{j}N}{c_{1}+c_{3}}+N^{i}\pi_{\phi}\partial_{i}\phi\nonumber \\
 & -\frac{4(c_{1}+c_{2})c_{3}h^{ij}h^{kl}\sqrt{h}\partial_{i}N\partial_{j}h_{kl}}{c_{1}+c_{3}}-\frac{4(c_{1}+c_{2})(c_{1}-c_{2}+2c_{3})h^{ij}\sqrt{h}\partial_{i}N\partial_{j}N}{(c_{1}+c_{3})N}\nonumber \\
 & +\frac{(c_{2}-\frac{c^{2}_{2}}{2c_{1}})\sqrt{h}\partial_{i}N^{j}\partial_{j}N^{i}}{N}+\frac{4(c_{1}+c_{2})N^{i}\sqrt{h}\partial_{i}N\partial_{j}N^{j}}{N^{2}}\nonumber \\
 & -\frac{(c_{1}+c_{2})(c^{2}_{1}+2c_{1}c_{3}-c_{2}c_{3})\sqrt{h}\partial_{i}N^{i}\partial_{j}N^{j}}{c_{1}(c_{1}+3c_{3})N}-\frac{(c_{1}+c_{2}-c_{3})h^{ij}h^{kl}N^{2}\pi_{i}\partial_{l}h_{jk}}{c_{1}+c_{3}}\nonumber \\
 & -\frac{4(c_{1}+c_{2})(c_{2}-2c_{3})h^{ij}h^{kl}\sqrt{h}\partial_{i}N\partial_{l}h_{jk}}{c_{1}+c_{3}}-\frac{c_{1}c_{3}h^{ij}h^{kl}h^{mn}N\sqrt{h}\partial_{k}h_{ij}\partial_{l}h_{mn}}{c_{1}+c_{3}}\nonumber \\
 & +\frac{(2c_{1}-\frac{c^{2}_{2}}{2c_{1}})h_{ik}h^{jl}\sqrt{h}\partial_{j}N^{i}\partial_{l}N^{k}}{N}-c_{2}h^{ij}h^{kl}h^{mn}N\sqrt{h}\partial_{l}h_{jn}\partial_{m}h_{ik}\nonumber \\
 & -c_{1}h^{ij}h^{kl}h^{mn}N\sqrt{h}\partial_{m}h_{ik}\partial_{n}h_{jl}-\frac{c_{2}h_{ik}\pi^{jk}\partial_{j}N^{i}}{c_{1}}\nonumber \\
 & +\frac{2(2c_{1}+c_{2})c_{3}h^{ij}h^{kl}h^{mn}N\sqrt{h}\partial_{k}h_{ij}\partial_{n}h_{lm}}{c_{1}+c_{3}}+\tfrac{1}{2}h^{ij}N\sqrt{h}\partial_{i}\phi\partial_{j}\phi-NV\left(\phi\right).
\end{align}
The total Hamiltonian is thus 
\begin{equation}
H^{(\mathrm{II})}_{\mathrm{T}}=H^{(\mathrm{II})}_{0}+\int\mathrm{d}^{3}y\,\lambda_{N}(\bm{y})\phi^{(\mathrm{II})}_{i}(\bm{y}),
\end{equation}
where $\lambda_{N}(\bm{y})$ is the Lagrange multiplier and the canonical
Hamiltonian is given by
\begin{equation}
H^{(\mathrm{II})}_{0}\left(t\right)\coloneqq\int\mathrm{d}^{3}x\mathcal{H}^{(\mathrm{II})}_{0}\left(t,\bm{x}\right).
\end{equation}

The time evolution of the primary constraint $\phi^{(\mathrm{II})}_{N}$
is determined by
\begin{equation}
\dot{\phi}^{(\mathrm{II})}_{N}(\bm{x})\approx\left[\phi^{(\mathrm{II})}_{N}(\bm{x}),H^{(\mathrm{II})}_{0}\right]+\int\mathrm{d}^{3}y\,\lambda_{N}(\bm{y})\left[\phi^{(\mathrm{II})}_{N}(\bm{x}),\phi^{(\mathrm{II})}_{N}(\bm{y})\right]=\left[\phi^{(\mathrm{II})}_{N}(\bm{x}),H^{(\mathrm{II})}_{0}\right],
\end{equation}
where $\left[\bullet,\bullet\right]$ denotes  the Poisson bracket,
and we have used the fact that $\left[\phi^{(\mathrm{II})}_{N}(\bm{x}),\phi^{(\mathrm{II})}_{N}(\bm{y})\right]\equiv0$.
It is easy to check that the Poisson bracket $\left[\phi^{(\mathrm{II})}_{N}\left(\bm{x}\right),H^{(\mathrm{II})}_{0}\right]$
does not vanish weakly (i.e., it is not proportional to $\phi^{(\mathrm{II})}_{N}$),
which thus yields a secondary constraint
\begin{equation}
\chi^{(\mathrm{II})}_{N}(\bm{x})\coloneqq\left[\phi^{(\mathrm{II})}_{N}\left(\bm{x}\right),H^{(\mathrm{II})}_{0}\right]\approx0.
\end{equation}
The concrete expression of $\chi_{N}$ is lengthy, and we do not present it
in the paper.

We can further consider the time evolution of $\chi_{N}$ by evaluating
\begin{equation}
\dot{\chi}^{(\mathrm{II})}_{N}(\bm{x})=\left[\chi^{(\mathrm{II})}_{N}(\bm{x}),H^{(\mathrm{II})}_{0}\right]+\int\mathrm{d}^{3}y\,\lambda_{N}(\bm{y})\left[\chi^{(\mathrm{II})}_{N}(\bm{x}),\phi^{(\mathrm{II})}_{N}(\bm{y})\right].
\end{equation}
In general, the Poisson bracket $\left[\chi^{(\mathrm{II})}_{N}(\bm{x}),\phi^{(\mathrm{II})}_{N}(\bm{y})\right]$
does not vanish weakly, which means that the consistency relation
of $\chi_{N}$ merely fixes the Lagrange multiplier $\lambda_{N}$
instead of generating a new constraint. The Dirac-Bergmann algorithm
thus stops. 

To summarize, we obtain one primary constraint and one secondary constraint,
both of which are second-class. The number of DOFs in case II is thus
calculated as
\begin{equation}
\#^{(\mathrm{II})}_{\mathrm{DOF}}=\frac{1}{2}\left(2\times\#_{\mathrm{var}}-2\times\#_{1\mathrm{st}}-\#_{2\mathrm{nd}}\right)=\frac{1}{2}\left(2\times11-2\right)=10,
\end{equation}
where $\#_{\mathrm{var}}$, $\#_{1\mathrm{st}}$, and $\#_{2\mathrm{nd}}$
stand for the numbers of configuration variables, first-class constraints,
and second-class constraints, respectively. This result is also consistent
with the counting of DOFs in the perturbative analysis in Sec. \ref{subsec:pert_case2}.

\subsection{Case V \label{subsec:Ham_case5}}

Now let us turn to case V with $c_{4}=-2c_{1}-c_{2}$. In this case,
the conjugate momenta are
\begin{align}
\pi^{ij}= & \sqrt{h}\bigg[-\frac{2c_{3}\dot{h}_{kl}h^{kl}h^{ij}}{N}-\frac{2c_{1}\dot{h}_{kl}h^{ik}h^{jl}}{N}-\frac{2\left(2c_{3}+c_{5}\right)\dot{N}h^{ij}}{N^{2}}\nonumber \\
 & +\frac{2c_{3}h^{lm}h^{ij}N^{k}\partial_{k}h_{lm}}{N}+\frac{2c_{1}h^{ik}h^{jl}N^{m}\partial_{m}h_{kl}}{N}+\frac{2\left(2c_{3}+c_{5}\right)h^{ij}N^{k}\partial_{k}N}{N^{2}}\nonumber \\
 & -\frac{c_{5}h^{ij}\partial_{k}N^{k}}{N}-\frac{c_{2}h^{jk}\partial_{k}N^{i}}{N}-\frac{c_{2}h^{ik}\partial_{k}N^{j}}{N}\bigg],\label{pi^ij_case5}
\end{align}
\begin{equation}
\pi_{i}=\sqrt{h}\left[\frac{2\left(2c_{1}+c_{2}\right)h^{jk}\partial_{k}h_{ij}}{N}-\frac{c_{5}h^{jk}\partial_{i}h_{jk}}{N}-\frac{2\left(2c_{2}+c_{5}\right)\partial_{i}N}{N^{2}}\right],\label{pi_i_case5}
\end{equation}
\begin{align}
\pi_{N}= & \sqrt{h}\bigg[\frac{8\left(c_{1}-c_{3}-c_{5}\right)\dot{N}}{N^{3}}-\frac{2\left(2c_{3}+c_{5}\right)\dot{h}_{ij}h^{ij}}{N^{2}}+\frac{2\left(2c_{3}+c_{5}\right)h^{ij}N^{k}\partial_{k}h_{ij}}{N^{2}}\nonumber \\
 & +\frac{8\left(-c_{1}+c_{3}+c_{5}\right)N^{i}\partial_{i}N}{N^{3}}+\frac{2\left(4c_{1}+2c_{2}-c_{5}\right)\partial_{i}N^{i}}{N^{2}}\bigg],\label{pi_N_case5}
\end{align}
and
\begin{equation}
\pi_{\phi}=\sqrt{h}\left(\frac{\dot{\phi}}{N}-\frac{N^{i}\partial_{i}\phi}{N}\right).
\end{equation}
Note in this case, the lapse function $N$ becomes dynamical. From
(\ref{pi_i_case5}), the velocity of $N^{i}$ cannot be solved, which
yields three primary constraints
\begin{equation}
\phi^{(\mathrm{V})}_{i}\coloneqq\pi_{i}-\sqrt{h}\left[\frac{2\left(2c_{1}+c_{2}\right)h^{jk}\partial_{k}h_{ij}}{N}-\frac{c_{5}h^{jk}\partial_{i}h_{jk}}{N}-\frac{2\left(2c_{2}+c_{5}\right)\partial_{i}N}{N^{2}}\right].
\end{equation}
After solving velocities $\dot{N}$ and $\dot{h}_{ij}$ from (\ref{pi_N_case5})
and (\ref{pi^ij_case5}), the canonical Hamiltonian density is given
by
\begin{align}
\mathcal{H}^{(\mathrm{V})}_{0}= & -N\sqrt{h}V(\phi)-\frac{N\pi^{ij}\pi_{ij}}{4c_{1}\sqrt{h}}+\frac{\left(4c_{1}c_{3}+c_{5}{}^{2}\right)N\pi^{2}}{4c_{1}\left[4\left(c_{1}-c_{3}-c_{5}\right)\left(3c_{3}+c_{1}\right)+3\left(2c_{3}+c_{5}\right)^{2}\right]\sqrt{h}}\nonumber \\
 & -\frac{\left(2c_{3}+c_{5}\right)N^{2}\pi\pi_{N}}{\left[8\left(c_{1}-c_{3}-c_{5}\right)\left(3c_{3}+c_{1}\right)+6\left(2c_{3}+c_{5}\right)^{2}\right]\sqrt{h}}+\frac{\left(c_{1}+3c_{3}\right)N^{3}\pi^{2}_{N}}{4\left[4\left(c_{1}-c_{3}-c_{5}\right)\left(3c_{3}+c_{1}\right)+3\left(2c_{3}+c_{5}\right)^{2}\right]\sqrt{h}}\nonumber \\
 & +\frac{\left(c_{1}+c_{2}\right)\left[c_{5}{}^{2}+2c_{1}\left(4c_{3}+c_{5}\right)\right]\pi\partial_{k}N^{k}}{c_{1}\left[4\left(c_{1}-c_{3}-c_{5}\right)\left(3c_{3}+c_{1}\right)+3\left(2c_{3}+c_{5}\right)^{2}\right]}+\frac{N\pi_{\phi}{}^{2}}{2\sqrt{h}}+N^{k}\pi^{ij}\partial_{k}h_{ij}+N^{k}\pi_{N}\partial_{k}N+N^{k}\pi_{\phi}\partial_{k}\phi\nonumber \\
 & -\frac{\bigl(8c_{1}{}^{2}+c_{1}(4c_{2}+24c_{3}-2c_{5})+3c_{5}{}^{2}+2c_{2}(8c_{3}+c_{5})\bigr)N\pi_{N}\partial_{k}N^{k}}{\left[8\left(c_{1}-c_{3}-c_{5}\right)\left(3c_{3}+c_{1}\right)+6\left(2c_{3}+c_{5}\right)^{2}\right]}\nonumber \\
 & -4c_{3}h^{ij}h^{kl}\sqrt{h}\partial_{i}N\partial_{j}h_{kl}-\frac{c_{5}h^{ij}N^{l}\sqrt{h}\partial_{l}N^{k}\partial_{k}h_{ij}}{N}-\frac{4(c_{1}+c_{3})h^{ij}\sqrt{h}\partial_{i}N\partial_{j}N}{N}\nonumber \\
 & -\frac{2(2c_{2}+c_{5})N^{i}\sqrt{h}\partial_{i}N^{j}\partial_{j}N}{N^{2}}-\frac{c_{2}h_{ij}\pi^{jk}\partial_{k}N^{i}}{c_{1}}+\frac{(c_{2}-\frac{c_{2}{}^{2}}{2c_{1}})\sqrt{h}\partial_{i}N^{j}\partial_{j}N^{i}}{N}\nonumber \\
 & +\frac{\left(c_{1}+c_{2}\right)\left\{ 8c_{1}{}^{3}+4c_{1}{}^{2}\left(c_{2}+8c_{3}\right)+c_{2}c_{5}{}^{2}+4c_{1}\left[c_{5}{}^{2}+c_{2}\left(6c_{3}+c_{5}\right)\right]\right\} \sqrt{h}\partial_{i}N^{i}\partial_{j}N^{j}}{c_{1}\left[4\left(c_{1}-c_{3}-c_{5}\right)\left(3c_{3}+c_{1}\right)+3\left(2c_{3}+c_{5}\right)^{2}\right]N}\nonumber \\
 & +\tfrac{1}{2}h^{ij}N\sqrt{h}\partial_{i}\phi\partial_{j}\phi-2c_{5}h^{ij}h^{kl}\sqrt{h}\partial_{i}N\partial_{l}h_{jk}\nonumber \\
 & +\frac{2\left(2c_{1}+c_{2}\right)h^{ij}N^{k}\sqrt{h}\partial_{k}N^{l}\partial_{j}h_{li}}{N}+\frac{\left(2c_{1}-\frac{c_{2}{}^{2}}{2c_{1}}\right)h_{ij}h^{kl}\sqrt{h}\partial_{k}N^{i}\partial_{l}N^{j}}{N}\nonumber \\
 & -c_{3}h^{ij}h^{kl}h^{mn}N\sqrt{h}\partial_{l}h_{mn}\partial_{k}h_{ij}+\left(2c_{1}+c_{2}\right)h^{ij}h^{kl}h^{mn}N\sqrt{h}\partial_{j}h_{ik}\partial_{n}h_{lm}\nonumber \\
 & -c_{5}h^{ij}h^{kl}h^{mn}N\sqrt{h}\partial_{k}h_{ij}\partial_{n}h_{lm}-c_{2}h^{ij}h^{kl}h^{mn}N\sqrt{h}\partial_{l}h_{jn}\partial_{m}h_{ik}\nonumber \\
 & -c_{1}h^{ij}h^{kl}h^{mn}N\sqrt{h}\partial_{n}h_{jl}\partial_{m}h_{ik}.
\end{align}
The total Hamiltonian is thus 
\begin{equation}
H^{(\mathrm{V})}_{\mathrm{T}}=H^{(\mathrm{V})}_{0}+\int\mathrm{d}^{3}y\,\lambda^{i}(\bm{y})\phi^{(\mathrm{V})}_{i}(\bm{y}),
\end{equation}
where $\lambda^{i}(\bm{y})$ are 3 Lagrange multipliers and the canonical
Hamiltonian is given by
\begin{equation}
H^{(\mathrm{V})}_{0}\left(t\right)\coloneqq\int\mathrm{d}^{3}x\mathcal{H}^{(\mathrm{V})}_{0}\left(t,\bm{x}\right).
\end{equation}

The time evolution of $\phi^{(\mathrm{V})}_{i}$ is given by
\begin{equation}
\dot{\phi}^{(\mathrm{V})}_{i}(\bm{x})\approx\left[\phi^{(\mathrm{V})}_{i}(\bm{x}),H_{0}\right]+\int\mathrm{d}^{3}y\,\lambda^{i}(\bm{y})\left[\phi^{(\mathrm{V})}_{i}(\bm{x}),\phi^{(\mathrm{V})}_{i}(\bm{y})\right].
\end{equation}
Since $\left[\phi^{(\mathrm{V})}_{i}(\bm{x}),\phi^{(\mathrm{V})}_{i}(\bm{y})\right]\equiv0$,
there will be three secondary constraints 
\begin{equation}
\chi^{(\mathrm{V})}_{i}(\bm{x})\coloneqq\left[\phi^{(\mathrm{V})}_{i}(\bm{x}),H^{(\mathrm{V})}_{0}\right]\approx0.
\end{equation}

The time evolution of the secondary constraints is determined by
\begin{equation}
\dot{\chi}^{(\mathrm{V})}_{i}(\bm{x})=\left[\chi^{(\mathrm{V})}_{i}(\bm{x}),H_{0}\right]+\int\mathrm{d}^{3}y\,\lambda^{i}(\bm{y})\left[\chi^{(\mathrm{V})}_{i}(\bm{x}),\phi^{(\mathrm{V})}_{i}(\bm{y})\right]\approx0.
\end{equation}
In general, the Poisson brackets among the primary and secondary constraints
$\left[\chi^{(\mathrm{V})}_{i}(\bm{x}),\phi^{(\mathrm{V})}_{i}(\bm{y})\right]$
are not vanishing, which implies that the consistency relation of
$\chi^{(\mathrm{V})}_{i}$ merely fixes the Lagrange multipliers $\lambda^{i}$
instead of generating new constraints\footnote{As being argued in \citep{DAmbrosio:2023asf}, in general the Lagrange
multipliers may satisfy partial differential equations (PDEs) with
spatial derivatives instead of purely algebraic equations. It is possible
that these PDEs do not yield unique solutions for the Lagrange multipliers,
which may imply the appearance of additional constraints. In this
work, we assume that all the Lagrange multipliers can have unique
solutions under appropriate spatial boundary conditions. }. To summarize, in case V, we have in total 6 constraints, which are
all second-class. The number of DOFs in case V is thus calculated
as 
\begin{equation}
\#^{(\mathrm{V})}_{\mathrm{DOF}}=\frac{1}{2}\left(2\times11-6\right)=8.
\end{equation}

Recall that in Sec. \ref{subsec:pert_case5}, we find only 6 DOFs
at the linear order in perturbations in a homogeneous and isotropic
background. This fact implies that there are 2 DOFs that manifest themselves
only at nonlinear orders in perturbation, or at linear order in an
inhomogeneous background. In the former case, this will result in
the strong coupling problem. This may be understood as resulting from
the enhanced degeneracy of the Hessian matrix on the FLRW background,
which renders these modes nondynamical. Nevertheless, these ``missing''
modes may reappear and become dynamical on less symmetric backgrounds, for instance on inhomogeneous backgrounds, or at nonlinear orders in perturbation theory.

\subsection{Case VI \label{subsec:Case-VI}}

Now let us turn to case VI with $c_{5}=2c_{1}$, $c_{4}=-2c_{1}-c_{2}$
and $c_{3}=-c_{1}$, where we have 2 free coefficients $c_{1}$ and
$c_{2}$. In this case, the velocity of the lapse function $\dot{N}$
drops out in the action. The conjugate momenta are 
\begin{align}
\pi^{ij}= & \sqrt{h}\bigg[2c_{1}\frac{\dot{h}_{kl}h^{kl}h^{ij}}{N}-2c_{1}\frac{\dot{h}_{kl}h^{ik}h^{jl}}{N}-2c_{1}\frac{h^{kl}h^{ij}N^{m}\partial_{m}h_{kl}}{N}\nonumber \\
 & +2c_{1}\frac{h^{ik}h^{jl}N^{m}\partial_{m}h_{kl}}{N}+2\left(c_{1}+2c_{2}\right)\frac{h^{ij}\partial_{k}N^{k}}{N}-c_{2}\frac{h^{jk}\partial_{k}N^{i}}{N}-c_{2}\frac{h^{ik}\partial_{k}N^{j}}{N}\bigg],
\end{align}
\begin{equation}
\pi_{i}=\sqrt{h}\frac{2}{N}\left(2c_{1}+c_{2}\right)\left(h^{jk}\partial_{j}h_{ik}-2h^{jk}\partial_{i}h_{jk}\right),
\end{equation}
\begin{equation}
\pi_{N}=0,
\end{equation}
and
\begin{equation}
\pi_{\phi}=\sqrt{h}\left(\frac{\dot{\phi}}{N}-\frac{N^{i}\partial_{i}\phi}{N}\right).
\end{equation}
We have four primary constraints 
\begin{equation}
\phi^{(\mathrm{VI})}_{N}\coloneqq\pi_{N}\approx0,
\end{equation}
and
\begin{equation}
\phi^{(\mathrm{VI})}_{i}=\pi_{i}-\sqrt{h}\frac{2}{N}\left(2c_{1}+c_{2}\right)\left(h^{jk}\partial_{j}h_{ik}-2h^{jk}\partial_{i}h_{jk}\right)\approx0.
\end{equation}

The canonical Hamiltonian density is 
\begin{align}
\mathcal{H}^{(\mathrm{VI})}_{0}= & \frac{N\pi^{2}}{8c_{1}\sqrt{h}}-\frac{N\pi_{ij}\pi^{ij}}{4c_{1}\sqrt{h}}+N^{i}\pi^{jk}\partial_{i}h_{jk}-\frac{(c_{1}+c_{2})h_{jk}\pi^{jk}\partial_{i}N^{i}}{2c_{1}}-\frac{c_{2}h_{ik}\pi^{jk}\partial_{j}N^{i}}{c_{1}}\nonumber \\
 & +4c_{1}h^{ij}h^{kl}\sqrt{h}\partial_{i}N\partial_{j}h_{kl}-\frac{2c_{1}h^{kl}N^{i}\sqrt{h}\partial_{i}N^{j}\partial_{j}h_{kl}}{N}-\frac{4(c_{1}+c_{2})N^{i}\sqrt{h}\partial_{i}N^{j}\partial_{j}N}{N^{2}}\nonumber \\
 & +\frac{(c_{2}-\frac{c^{2}_{2}}{2c_{1}})\sqrt{h}\partial_{i}N^{j}\partial_{j}N^{i}}{N}-\frac{4(c_{1}+c_{2})h^{kl}N^{i}\sqrt{h}\partial_{i}h_{kl}\partial_{j}N^{j}}{N}+\frac{4(c_{1}+c_{2})N^{i}\sqrt{h}\partial_{i}N\partial_{j}N^{j}}{N^{2}}\nonumber \\
 & +\frac{(-\tfrac{1}{2}c_{1}+4c_{2}+\frac{9c^{2}_{2}}{2c_{1}})\sqrt{h}\partial_{i}N^{i}\partial_{j}N^{j}}{N}-4c_{1}h^{ij}h^{kl}\sqrt{h}\partial_{i}N\partial_{l}h_{jk}+\frac{2(2c_{1}+c_{2})h^{kl}N^{i}\sqrt{h}\partial_{i}N^{j}\partial_{l}h_{jk}}{N}\nonumber \\
 & +c_{1}h^{ij}h^{kl}h^{mn}N\sqrt{h}\partial_{k}h_{ij}\partial_{l}h_{mn}+\frac{(2c_{1}-\frac{c^{2}_{2}}{2c_{1}})h_{ik}h^{jl}\sqrt{h}\partial_{j}N^{i}\partial_{l}N^{k}}{N}-c_{2}h^{ij}h^{kl}h^{mn}N\sqrt{h}\partial_{l}h_{jn}\partial_{m}h_{ik}\nonumber \\
 & -c_{1}h^{ij}h^{kl}h^{mn}N\sqrt{h}\partial_{m}h_{ik}\partial_{n}h_{jl}+(2c_{1}+c_{2})h^{ij}h^{kl}h^{mn}N\sqrt{h}\partial_{j}h_{ik}\partial_{n}h_{lm}-2c_{1}h^{ij}h^{kl}h^{mn}N\sqrt{h}\partial_{k}h_{ij}\partial_{n}h_{lm}\nonumber \\
 & +\frac{N\pi^{2}_{\varphi}}{2\sqrt{h}}+N^{i}\pi_{\varphi}\partial_{i}\varphi+\tfrac{1}{2}\sqrt{h}Nh^{ij}\partial_{i}\varphi\partial_{j}\varphi-NV\left(\phi\right).
\end{align}
For the sake of brevity, we denote the set of primary constraints
as
\begin{equation}
\left\{ \phi_{a}\right\} \equiv\left\{ \phi^{(\mathrm{VI})}_{N},\phi^{(\mathrm{VI})}_{i}\right\} ,\quad a=1,\cdots,4.
\end{equation}
The total Hamiltonian is thus defined by
\begin{equation}
H^{(\mathrm{VI})}_{\mathrm{T}}=H^{(\mathrm{VI})}_{0}+\int\mathrm{d}^{3}x\,\lambda^{a}(\bm{x})\phi^{(\mathrm{VI})}_{a}(\bm{x}),
\end{equation}
where $H^{(\mathrm{VI})}_{0}$ is the canonical Hamiltonian given
by
\begin{equation}
H^{(\mathrm{VI})}_{0}\left(t\right)=\int\mathrm{d^{3}}x\,\mathcal{H}^{(\mathrm{VI})}_{0}\left(t,\bm{x}\right),
\end{equation}
and the set of Lagrange multipliers is
\begin{equation}
\left\{ \lambda^{a}\right\} \equiv\left\{ \lambda_{N},\lambda^{i}\right\} ,\quad a=1,\cdots,4.
\end{equation}

The consistency relation of the primary constraints is thus
\begin{equation}
\dot{\phi}^{(\mathrm{VI})}_{a}(\bm{x})\approx\left[\phi^{(\mathrm{VI})}_{a}(\bm{x}),H^{(\mathrm{VI})}_{0}\right]+\int\mathrm{d}^{3}y\,D_{ab}(\bm{x},\bm{y})\lambda^{b}(\bm{y})\approx0,\label{consrel_case6}
\end{equation}
where we denote the Poisson brackets among the primary constraints
\begin{equation}
D_{ab}(\bm{x},\bm{y})\coloneqq\left[\phi^{(\mathrm{VI})}_{a}(\bm{x}),\phi^{(\mathrm{VI})}_{b}(\bm{y})\right]\label{Dab_case6}
\end{equation}
for later convenience. The Poisson brackets among the primary constraints
are given by
\begin{equation}
\left[\phi^{(\mathrm{VI})}_{N}(\bm{x}),\phi^{(\mathrm{VI})}_{N}(\bm{y})\right]=\left[\phi^{(\mathrm{VI})}_{i}(\bm{x}),\phi^{(\mathrm{VI})}_{j}(\bm{y})\right]=0
\end{equation}
and
\begin{equation}
\left[\phi^{(\mathrm{VI})}_{i}(\bm{x}),\phi^{(\mathrm{VI})}_{N}(\bm{y})\right]=\frac{2}{N^{2}}\left(2c_{1}+c_{2}\right)\left(h^{jk}\partial_{j}h_{ik}-2h^{jk}\partial_{i}h_{jk}\right)\delta^{3}\left(\boldsymbol{x}-\boldsymbol{y}\right).\label{PB_phi_i_phi_N_case6}
\end{equation}

From (\ref{PB_phi_i_phi_N_case6}), according to whether $2c_{1}+c_{2}=0$
or not, we can discuss two subcases. In the special case, we have
$2c_{1}+c_{2}=0$ and thus $\left[\phi^{(\mathrm{VI})}_{i}(\bm{x}),\phi^{(\mathrm{VI})}_{N}(\bm{y})\right]=0$.
This case is nothing but STEGR. The Hamiltonian constraint analysis
has been performed in \citep{DAmbrosio:2020nqu} and here we briefly
mention the main results. In this case, the primary constraints reduce
to
\begin{equation}
\phi^{(\mathrm{VI})}_{N}=\pi_{N}\approx0,\quad\phi^{(\mathrm{VI})}_{i}=\pi_{i}\approx0,
\end{equation}
and the consistency relations (\ref{consrel_case6}) yield four secondary
constraints, 
\begin{align}
\mathcal{C}_{0}\left(\boldsymbol{x}\right) & \coloneqq\left[\phi^{(\mathrm{VI})}_{N}\left(\boldsymbol{x}\right),H^{(\mathrm{VI})}_{0}\right]\approx-\sqrt{h}\left[\overset{3}{\mathcal{Q}}-h^{ij}h^{kl}\mathrm{D}_{i}\left(Q_{jkl}-Q_{kjl}\right)-\frac{1}{h}\left(\pi_{ij}\pi^{ij}-\frac{1}{2}\pi^{2}\right)\right],\\
\mathcal{C}_{i}\left(\boldsymbol{x}\right) & \coloneqq\left[\phi^{(\mathrm{VI})}_{i}\left(\boldsymbol{x}\right),H^{(\mathrm{VI})}_{0}\right]\approx-2\mathrm{D}_{j}\pi^{j}_{\phantom{j}i}.
\end{align}
Here $\mathrm{D}_{i}$ is the spatial covariant derivative adapted
to the spatial metric $h_{ij}$. The Hamiltonian density can be brought
into the form
\begin{equation}
\mathcal{H}^{(\mathrm{VI})}_{0}=N\mathcal{C}_{0}+N^{i}\mathcal{C}_{i}.
\end{equation}
By introducing the smeared functionals \citep{DAmbrosio:2020nqu}
\begin{align}
C_{S}\left[\alpha\right] & \coloneqq\int\mathrm{d}x^{3}\,\alpha\left(\boldsymbol{x}\right)\mathcal{C}_{0}\left(\boldsymbol{x}\right),\\
C_{V}\left[\vec{\beta}\right] & \coloneqq\int\mathrm{d}x^{3}\,\beta^{i}\left(\boldsymbol{x}\right)\mathcal{C}_{i}\left(\boldsymbol{x}\right),
\end{align}
where $\alpha$ and $\beta^{i}$ are some smearing functions, after
some manipulations, one can show the Poisson brackets 
\begin{align}
\left[C_{S}\left[\alpha\right],C_{V}\left[\vec{\beta}\right]\right] & =-C_{S}\left[\mathcal{L}_{\vec{\beta}}\alpha\right],\\
\left[C_{V}\left[\vec{\beta}_{1}\right],C_{V}\left[\vec{\beta}_{2}\right]\right] & =C_{V}\left[\left[\vec{\beta}_{1},\vec{\beta}_{2}\right]\right],\\
\left[C_{S}\left[\alpha_{1}\right],C_{S}\left[\alpha_{2}\right]\right] & =C_{V}\left[\left(\alpha_{1}\partial^{i}\alpha_{2}-\alpha_{2}\partial^{i}\alpha_{1}\right)\partial_{i}\right],
\end{align}
which are all vanishing weakly. As a result, the consistency relation
of the secondary constraints does not yield further constraints. Thus,
we have in total four first-class primary constraints and four first-class
secondary constraints. Therefore, the number of DOFs in this special
case is calculated as
\begin{equation}
\#^{(\mathrm{VI})}_{\mathrm{DOF}}=\frac{1}{2}\left(22-2\times8\right)=3.
\end{equation}

Generally, $2c_{1}+c_{2}\neq0$, and thus $\left[\phi^{(\mathrm{VI})}_{i}(\bm{x}),\phi^{(\mathrm{VI})}_{N}(\bm{y})\right]\neq0$.
For the sake of brevity, we denote
\begin{equation}
\varDelta_{i}(\bm{x},\bm{y})\equiv\left[\phi_{N}(\bm{x}),\phi_{i}(\bm{y})\right],\quad i=1,2,3,
\end{equation}
and define its ``inverse'' by\footnote{Alternatively, one may define its inverse through $\int\mathrm{d}^{3}x'\varDelta_{i}(\bm{x},\bm{x}')\tilde{\varDelta}^{-1}_{i}(\bm{x}',\bm{y})=\delta^{3}(\bm{x}-\bm{y})$.
It is easy to show that $\varDelta^{-1}_{i}(\bm{x},\bm{y})\equiv\tilde{\varDelta}^{-1}_{i}(\bm{x},\bm{y})$.}
\begin{equation}
\int\mathrm{d}^{3}x'\varDelta^{-1}_{i}(\bm{x},\bm{x}')\varDelta_{i}(\bm{x}',\bm{y})=\delta^{3}(\bm{x}-\bm{y}),\quad i=1,2,3.
\end{equation}
Thus $D_{ab}(\bm{x},\bm{y})$ defined in (\ref{Dab_case6}) takes
the form
\begin{equation}
D_{ab}(\bm{x},\bm{y})=\left(\begin{array}{cccc}
0 & \varDelta_{1}(\bm{x},\bm{y}) & \varDelta_{2}(\bm{x},\bm{y}) & \varDelta_{3}(\bm{x},\bm{y})\\
-\varDelta_{1}(\bm{y},\bm{x}) & 0 & 0 & 0\\
-\varDelta_{2}(\bm{y},\bm{x}) & 0 & 0 & 0\\
-\varDelta_{3}(\bm{y},\bm{x}) & 0 & 0 & 0
\end{array}\right).
\end{equation}

Next we will show that the rank of $D_{ab}(\bm{x},\bm{y})$ is 2,
i.e., $D_{ab}(\bm{x},\bm{y})$ has 2 null eigenvectors satisfying
\begin{equation}
\int\mathrm{d}^{3}y\,D_{ab}(\bm{x},\bm{y})V^{b}_{(\alpha)}(\bm{y})=0,\quad\alpha=1,2.\label{Dab_eigen}
\end{equation}
To this end, we denote
\begin{equation}
V^{b}_{(\alpha)}(\bm{y})=\left(\begin{array}{c}
V^{0}(\bm{y})\\
V^{1}(\bm{y})\\
V^{2}(\bm{y})\\
V^{3}(\bm{y})
\end{array}\right),
\end{equation}
thus (\ref{Dab_eigen}) yields
\begin{equation}
\int\mathrm{d}^{3}y\left(\begin{array}{c}
\varDelta_{1}(\bm{x},\bm{y})V^{1}(\bm{y})+\varDelta_{2}(\bm{x},\bm{y})V^{2}(\bm{y})+\varDelta_{3}(\bm{x},\bm{y})V^{3}(\bm{y})\\
-\varDelta_{1}(\bm{y},\bm{x})V^{0}(\bm{y})\\
-\varDelta_{2}(\bm{y},\bm{x})V^{0}(\bm{y})\\
-\varDelta_{3}(\bm{y},\bm{x})V^{0}(\bm{y})
\end{array}\right)=0.
\end{equation}
Therefore, we must have
\begin{equation}
V^{0}\equiv0
\end{equation}
and
\begin{equation}
\int\mathrm{d}^{3}y\left(\varDelta_{1}(\bm{x},\bm{y})V^{1}(\bm{y})+\varDelta_{2}(\bm{x},\bm{y})V^{2}(\bm{y})+\varDelta_{3}(\bm{x},\bm{y})V^{3}(\bm{y})\right)=0.\label{case6_cond}
\end{equation}

Equation (\ref{case6_cond}) is for 3 functions $V^{i}(\bm{y})$,
which has two solutions with one arbitrary (unfixed) function. Without
loss of generality, we denote the arbitrary function as $V^{1}(\bm{y})\equiv V(\bm{y})\neq0$.
The first solution corresponds to $V^{2}\neq0$ and $V^{3}=0$. Thus,
Eq. (\ref{case6_cond}) implies
\begin{equation}
\int\mathrm{d}^{3}y\varDelta_{2}(\bm{x}',\bm{y})V^{2}(\bm{y})=-\int\mathrm{d}^{3}y\varDelta_{1}(\bm{x}',\bm{y})V(\bm{y}),
\end{equation}
multiplying both sides by $\int\mathrm{d}^{3}x'\varDelta^{-1}_{2}(\bm{x},\bm{x}')$
yields
\begin{equation}
V^{2}(\bm{x})=-\int\mathrm{d}^{3}x'\int\mathrm{d}^{3}y\varDelta^{-1}_{2}(\bm{x},\bm{x}')\varDelta_{1}(\bm{x}',\bm{y})V(\bm{y}).
\end{equation}
Similarly, the other solution corresponds to setting $V^{2}=0$ and
$V^{3}\neq0$, which gives
\begin{equation}
V^{3}(\bm{x})=-\int\mathrm{d}^{3}x'\int\mathrm{d}^{3}y\varDelta^{-1}_{3}(\bm{x},\bm{x}')\varDelta_{1}(\bm{x}',\bm{y})V(\bm{y}).
\end{equation}

The 2 null eigenvectors are thus
\begin{equation}
V^{b}_{(1)}(\bm{x})=\left(\begin{array}{c}
0\\
\int\mathrm{d}^{3}y\delta^{3}(\bm{x}-\bm{y})\\
-\int\mathrm{d}^{3}x'\int\mathrm{d}^{3}y\varDelta^{-1}_{2}(\bm{x},\bm{x}')\varDelta_{1}(\bm{x}',\bm{y})\\
0
\end{array}\right)V(\bm{y}),\label{nev_case6_1}
\end{equation}
and
\begin{equation}
V^{b}_{(2)}(\bm{x})=\left(\begin{array}{c}
0\\
\int\mathrm{d}^{3}y\delta^{3}(\bm{x}-\bm{y})\\
0\\
-\int\mathrm{d}^{3}x'\int\mathrm{d}^{3}y\varDelta^{-1}_{3}(\bm{x},\bm{x}')\varDelta_{1}(\bm{x}',\bm{y})
\end{array}\right)V(\bm{y}).\label{nev_case6_2}
\end{equation}

By multiplying both sides of the consistency relation (\ref{consrel_case6})
by $\int\mathrm{d}^{3}xV^{a}_{(\alpha)}(\bm{x})$, we get
\begin{equation}
\int\mathrm{d}^{3}xV^{a}_{(\alpha)}(\bm{x})\left[\phi^{(\mathrm{VI})}_{a}(\bm{x}),H^{(\mathrm{VI})}_{0}\right]\approx-\int\mathrm{d}^{3}xV^{a}_{(\alpha)}(\bm{x})\int\mathrm{d}^{3}y\,D_{ab}(\bm{x},\bm{y})\lambda^{b}(\bm{y})=0,\quad\alpha=1,2,
\end{equation}
which yields two secondary constraints
\begin{equation}
X^{(\mathrm{VI})}_{\alpha}=\int\mathrm{d}^{3}xV^{a}_{(\alpha)}(\bm{x})\left[\phi^{(\mathrm{VI})}_{a}(\bm{x}),H^{(\mathrm{VI})}_{0}\right].
\end{equation}
Plugging the concrete solutions of (\ref{nev_case6_1}) and (\ref{nev_case6_2}),
we get
\begin{align}
X^{(\mathrm{VI})}_{1}= & \int\mathrm{d}^{3}x\int\mathrm{d}^{3}y\,\delta^{3}(\bm{x}-\bm{y})V(\bm{y})\left[\phi^{(\mathrm{VI})}_{1}(\bm{x}),H^{(\mathrm{VI})}_{0}\right]\nonumber \\
 & -\int\mathrm{d}^{3}x\int\mathrm{d}^{3}y\int\mathrm{d}^{3}x'\varDelta^{-1}_{2}(\bm{x},\bm{x}')\varDelta_{1}(\bm{x}',\bm{y})V(\bm{y})\left[\phi^{(\mathrm{VI})}_{2}(\bm{x}),H^{(\mathrm{VI})}_{0}\right].
\end{align}
Since $V(\bm{y})$ is an arbitrary function, we may define the local
constraint
\begin{equation}
X^{(\mathrm{VI})}_{\alpha}=\int\mathrm{d}^{3}y\,\chi^{(\mathrm{VI})}_{\alpha}(\bm{y})V(\bm{y}),\quad\alpha=1,2,
\end{equation}
where
\begin{align}
\chi^{(\mathrm{VI})}_{1}(\bm{y})= & \left[\phi^{(\mathrm{VI})}_{1}(\bm{y}),H^{(\mathrm{VI})}_{0}\right]\nonumber \\
 & -\int\mathrm{d}^{3}x\int\mathrm{d}^{3}x'\varDelta^{-1}_{2}(\bm{x},\bm{x}')\varDelta_{1}(\bm{x}',\bm{y})\left[\phi^{(\mathrm{VI})}_{2}(\bm{x}),H^{(\mathrm{VI})}_{0}\right].
\end{align}
Similarly, the other constraint is
\begin{align}
\chi^{(\mathrm{VI})}_{2}(\bm{y})= & \left[\phi^{(\mathrm{VI})}_{1}(\bm{y}),H^{(\mathrm{VI})}_{0}\right]\nonumber \\
 & -\int\mathrm{d}^{3}x\int\mathrm{d}^{3}x'\varDelta^{-1}_{3}(\bm{x},\bm{x}')\varDelta_{1}(\bm{x}',\bm{y})\left[\phi^{(\mathrm{VI})}_{3}(\bm{x}),H^{(\mathrm{VI})}_{0}\right].
\end{align}
The explicit expressions for these two secondary constraints are extremely
complicated, and we do not present them in the paper. 

The time evolution of the secondary constraints is given by
\begin{equation}
\dot{\chi}^{(\mathrm{VI})}_{\alpha}(\bm{x})\approx\left[\chi^{(\mathrm{VI})}_{\alpha}(\bm{x}),H^{(\mathrm{VI})}_{0}\right]+\int\mathrm{d}^{3}y\,\left[\chi^{(\mathrm{VI})}_{\alpha}(\bm{x}),\phi^{(\mathrm{VI})}_{b}(\bm{y})\right]\lambda^{b}(\bm{y})\approx0.
\end{equation}
It can be found that without further conditions, the consistency relations
of the secondary constraints merely fix the Lagrange multipliers instead
of generating new constraints. As a result, we have in total 6 constraints,
which are all second-class. Therefore, the number of DOFs in case
VI is generally given by
\begin{equation}
\#^{(\mathrm{VI})}_{\mathrm{DOFs}}=\frac{1}{2}\left(2\times11-6\right)=8.
\end{equation}

Recall that in the previous section, we have calculated the linear
perturbations of this case and found that there are 5 DOFs at the
linear order in perturbations. Again, this fact implies that there
are 3 DOFs that manifest themselves only at nonlinear order in perturbations,
or at linear order in an inhomogeneous background, which may signal
the strong coupling problem. 

\subsubsection{$f(Q)$ theory \label{subsec:fQ-theory}}

At this point, recall that the $f(Q)$ theory in the presence of the kinetic
terms of the scalar field can be regarded as a special case of case
VI (see Table \ref{tab:cases_phy}). It is thus interesting to see
the implication on the number of DOFs of $f(Q)$ theory. 
To apply the results of the above analysis, we need to turn
off the kinetic term of the scalar field. This yields an additional
primary constraint,
\begin{equation}
\phi^{(\mathrm{VI})}_{\phi}\coloneqq\pi_{\phi}\approx0.
\end{equation}
Its Poisson brackets with all the primary constraints are given by
\begin{equation}
\left[\phi^{(\mathrm{VI})}_{\phi}(\bm{x}),\phi^{(\mathrm{VI})}_{\phi}(\bm{y})\right]=\left[\phi^{(\mathrm{VI})}_{\phi}(\bm{x}),\phi^{(\mathrm{VI})}_{N}(\bm{y})\right]=0
\end{equation}
and
\begin{equation}
\left[\phi^{(\mathrm{VI})}_{\phi}(\bm{x}),\phi^{(\mathrm{VI})}_{i}(\bm{y})\right]=-\frac{2}{N}\left(2c_{1}'+c_{2}'\right)\left(h^{jk}\partial_{j}h_{ik}-2h^{jk}\partial_{i}h_{jk}\right)\delta^{3}\left(\boldsymbol{x}-\boldsymbol{y}\right),
\end{equation}
where $c_{i}'\equiv c_{i,\phi}\left(\phi\right)$, for $i=1,2,3,4,5$.

Now we have a set of primary constraints
\begin{equation}
\left\{ \phi_{a}\right\} \equiv\left\{ \phi^{(\mathrm{VI})}_{N},\phi^{(\mathrm{VI})}_{i},\phi^{(\mathrm{VI})}_{\phi}\right\} ,\quad a=1,\cdots,5.
\end{equation}
If we consider a linear combination of $\phi^{(\mathrm{VI})}_{\phi}$
and $\phi^{(\mathrm{VI})}_{N}$, which is defined as
\begin{equation}
\tilde{\phi}^{(\mathrm{VI})}\coloneqq\left(2c_{1}+c_{2}\right)\phi^{(\mathrm{VI})}_{\phi}-\left(2c_{1}'+c_{2}'\right)N\phi^{(\mathrm{VI})}_{N},
\end{equation}
then we find its Poisson brackets where all the primary constraints
vanish,
\begin{equation}
\left[\tilde{\phi}^{(\mathrm{VI})}(\bm{x}),\phi_{a}(\bm{y})\right]=0,\quad a=1,\cdots,5.
\end{equation}

We thus regard $\left\{ \phi^{(\mathrm{VI})}_{N},\phi^{(\mathrm{VI})}_{i},\tilde{\phi}^{(\mathrm{VI})}\right\} $
as our new complete set of primary constraints. Therefore, it is evident
that when considering the consistency relation of the primary constraints,
this new primary constraint (due to the absence of the kinetic term
of the scalar field) will lead to an additional secondary constraint,
which is given by
\begin{equation}
\tilde{\chi}^{(\mathrm{VI})}\coloneqq\left[\tilde{\phi}^{(\mathrm{VI})}(\bm{x}),H^{(\mathrm{VI})}_{0}\right]\approx0.
\end{equation}
As a result, in the absence of the kinetic term of the scalar field,
this theory will yield two extra constraints $\tilde{\phi}^{(\mathrm{VI})}$
and $\tilde{\chi}^{(\mathrm{VI})}$. It can be checked that both are
second-class constraints. Consequently, it can be concluded that the
number of DOFs of the $f\left(Q\right)$ theory is given by
\begin{equation}
\#^{(f\left(Q\right))}_{\mathrm{DOFs}}=\frac{1}{2}\left(2\times11-8\right)=7.
\end{equation}
This conclusion is consistent with several recent analyses on the
propagating DOFs in $f(Q)$ gravity \citep{DAmbrosio:2023asf,Tomonari:2023wcs,Gomes:2023tur,Heisenberg:2025fxc},
while also helping to clarify how that result emerges within the present
constraint analysis .

\section{Conclusion}

In this work we determined the number of propagating
DOFs in the QSN theory, whose Lagrangian
consists of a general linear combination of the five independent quadratic
nonmetricity invariants with coefficients promoted to functions of
a dynamical scalar field. This framework contains the
STEGR and the widely studied $f(Q)$
model as special cases. To obtain robust DOF counts, we combined two
complementary approaches: linear cosmological perturbations around
an FLRW background and a fully nonperturbative Dirac-Bergmann Hamiltonian
constraint analysis.

In Sec. \ref{sec:pricons}, after performing the ADM decomposition
of the QSN action, we classified the theory into distinct cases according
to the number and structure of primary constraints. The classification
is summarized in Tables \ref{tab:cases_phy} and \ref{tab:cases_unphy},
and naturally separates into two classes. The cases in Table \ref{tab:cases_phy}
admit propagating tensor modes and are therefore phenomenologically
viable in the gravitational wave sector. The cases in Table \ref{tab:cases_unphy}
lack propagating tensor modes and are thus phenomenologically disfavored
from this perspective.

In Sec. \ref{sec:pert}, we then investigated linear cosmological
perturbations for the physical class. We found that cases III, IV,
and VII do not admit a consistent cosmological background, and therefore
focused on cases II, V, and VI. At linear order in cosmological perturbations,
these three cases propagate 10, 6, and 5 DOFs, respectively.

In Sec. \ref{sec:sec_cons}, we carried out the Hamiltonian constraint
analysis for cases II, V, and VI by tracking the time evolution of
the primary constraints obtained in Sec. \ref{sec:pricons}. For case
II, the preservation of the primary constraint generates a single
secondary constraint $\chi^{(\mathrm{II})}_{N}$ and no tertiary constraints.
Hence, there are two constraints in total, both of which are second-class,
leading to 10 propagating DOFs. This agrees with the perturbative
result, in the sense that all 10 DOFs are already visible at linear
order. For case V, the consistency of the three primary constraints
$\phi^{(\mathrm{V})}_{i}$ yields three secondary constraints $\chi^{(\mathrm{V})}_{i}$.
For case VI, there are four primary constraints $\phi^{(\mathrm{VI})}_{N}$
and $\phi^{(\mathrm{VI})}_{i}$, while their consistency conditions
generate only two secondary constraints $\chi^{(\mathrm{VI})}_{1}$
and $\chi^{(\mathrm{VI})}_{2}$. In both cases V and VI, the Dirac
matrices constructed from the Poisson brackets among the primary and
secondary constraints are generically nondegenerate. Therefore, without
imposing further conditions, no tertiary constraints arise. Consequently,
both cases V and VI possess 6 second-class constraints in total and
propagate 8 DOFs at the nonperturbative level. As a consistency check,
we verified that STEGR (with a dynamical scalar field) is contained
in case VI as a special limit and propagates 3 DOFs.

Finally, comparing the two counting methods reveals a nontrivial mismatch
in cases V and VI. The Hamiltonian analysis gives 8 DOFs for both
cases, while only 6 and 5 modes, respectively, appear at linear order
on an FLRW background. This mismatch indicates that some of the nonperturbative
DOFs have vanishing quadratic kinetic terms on an FLRW background,
so they are not captured by the linearized cosmological perturbations.
These modes can become dynamical either beyond linear order around
a cosmological background or already at linear order on less symmetric
backgrounds (e.g., inhomogeneous configurations). In this sense, cases
V and VI exhibit a potential strong-coupling problem on cosmological
backgrounds. We leave a dedicated determination of the associated
strong-coupling scale to future work.
\begin{acknowledgments}
X. G. is supported by the National Natural Science Foundation of China
(NSFC) under Grants No. 12475068 and No. 11975020 and the Guangdong
Basic and Applied Basic Research Foundation under Grant No. 2025A1515012977.
\end{acknowledgments}

\section*{Data Availability}
No data were created or analyzed in this study.

\appendix

\section{ADM decomposition of the nonmetricity tensor \label{app:decnm}}

In this appendix, we show the details in deriving the ADM decomposition
of quadratic nonmetricity scalars in the coincident gauge. 

We denote $n^{\mu}$ as the normal vector to the spatial hypersurfaces,
which is normalized and timelike, i.e., $n^{\mu}n_{\mu}=-1$. The
spacetime therefore can be foliated by a family of spacelike hypersurfaces.
The induced metric on the hypersurface is defined by
\begin{equation}
h_{\mu\nu}=g_{\mu\nu}+n_{\mu}n_{\nu}.
\end{equation}

The nonmetricity tensor $Q_{\rho\mu\nu}$ can be decomposed as
\begin{align}
Q_{\rho\mu\nu}= & Q_{\hat{\rho}\hat{\mu}\hat{\nu}}-Q_{\hat{\rho}\hat{\mu}\bm{n}}n_{\nu}-Q_{\hat{\rho}\hat{\nu}\bm{n}}n_{\mu}+Q_{\hat{\rho}\bm{n}\bm{n}}n_{\mu}n_{\nu}\nonumber \\
 & -n_{\rho}Q_{\bm{n}\hat{\mu}\hat{\nu}}+Q_{\bm{n}\bm{n}\hat{\mu}}n_{\rho}n_{\nu}+Q_{\bm{n}\bm{n}\hat{\nu}}n_{\rho}n_{\mu}-n_{\rho}n_{\mu}n_{\nu}Q_{\bm{n}\bm{n}\bm{n}},
\end{align}
where\footnote{Here we follow the notation in \citep{Deruelle:2009zk}, where an
index replaced by $\bm{n}$ denotes contraction with the normal vector,
and a hatted-index denotes contraction with the spatial metric.}
\begin{align}
Q_{\bm{n}\bm{n}\bm{n}} & \coloneqq n^{\rho}n^{\mu}n^{\nu}Q_{\rho\mu\nu},\\
Q_{\bm{n}\bm{n}\hat{\nu}} & \coloneqq n^{\rho}n^{\mu}h^{\phantom{\nu}\nu^{\prime}}_{\nu}Q_{\rho\mu\nu^{\prime}},\\
Q_{\hat{\rho}\bm{n}\bm{n}} & \coloneqq h^{\phantom{\rho}\rho^{\prime}}_{\rho}n^{\mu}n^{\nu}Q_{\rho\mu\nu},\\
Q_{\hat{\rho}\hat{\mu}\bm{n}} & \coloneqq h^{\phantom{\rho}\rho^{\prime}}_{\rho}h^{\phantom{\mu}\mu^{\prime}}_{\mu}n^{\nu}Q_{\rho\mu\nu},\\
Q_{\bm{n}\hat{\mu}\hat{\nu}} & \coloneqq n^{\rho}h^{\phantom{\mu}\mu^{\prime}}_{\mu}h^{\phantom{\nu}\nu^{\prime}}_{\nu}Q_{\rho\mu^{\prime}\nu^{\prime}},\\
Q_{\hat{\rho}\hat{\mu}\hat{\nu}} & \coloneqq h^{\rho'}_{\rho}h^{\mu'}_{\mu}h^{\nu'}_{\nu}Q_{\rho'\mu'\nu'}
\end{align}
are all spatial tensors; i.e., they have vanishing contractions with the
normal vector.

By choosing the coincident gauge, $Q_{\rho\mu\nu}=\partial_{\rho}g_{\mu\nu}$.
In ADM coordinates, only the spatial components of the above tensors
will contribute, which are given by
\begin{align}
Q_{\bm{n}\bm{n}\bm{n}} & =\frac{2}{N^{2}}\left(N^{i}\partial_{i}N-\dot{N}\right),\\
Q_{\bm{n}\bm{n}k} & =\frac{h_{km}}{N^{2}}\left(\dot{N}^{m}-N^{i}\partial_{i}N^{m}\right),\\
Q_{i\bm{n}\bm{n}} & =-2\frac{\partial_{i}N}{N},\\
Q_{ik\bm{n}} & =\frac{h_{jk}\partial_{i}N^{j}}{N},\\
Q_{\bm{n}jk} & =2K_{jk}+\frac{h_{km}\partial_{j}N^{m}+h_{mj}\partial_{k}N^{m}}{N},\\
Q_{ijk} & =\partial_{i}h_{jk}.
\end{align}
Please note that after choosing the coincident gauge, we have no spatial
covariance any more, and thus the above components are not spatial
tensors.

According to these equations, after tedious calculations, we can obtain
the ADM decomposition of the 5 quadratic nonmetricity scalars in (\ref{QSN_action}):
\begin{align}
Q_{\rho\mu\nu}Q^{\rho\mu\nu}= & -4h^{ij}h^{kl}K_{ik}K_{jl}-\frac{4\dot{N}^{2}}{N^{4}}+\frac{2h_{ij}\dot{N}^{i}\dot{N}^{j}}{N^{4}}+\frac{8\dot{N}N^{i}\partial_{i}N}{N^{4}}-\frac{4h_{ij}N^{k}\dot{N}^{j}\partial_{k}N^{i}}{N^{4}}\nonumber \\
 & +h^{ij}h^{kl}h^{mn}\partial_{k}h_{im}\partial_{l}h_{jn}+\frac{4h^{ij}\partial_{i}N\partial_{j}N}{N^{2}}-\frac{4N^{i}N^{j}\partial_{i}N\partial_{j}N}{N^{4}}\nonumber \\
 & -\frac{8h^{ij}K_{jk}\partial_{i}N^{k}}{N}+\frac{2h_{ij}N^{k}N^{l}\partial_{k}N^{i}\partial_{l}N^{j}}{N^{4}}-\frac{2\partial_{i}N^{j}\partial_{j}N^{i}}{N^{2}}-\frac{4h_{ij}h^{kl}\partial_{k}N^{i}\partial_{l}N^{j}}{N^{2}},
\end{align}
\begin{align}
Q_{\rho\mu\nu}Q^{\mu\nu\rho}= & -\frac{4\dot{N}^{2}}{N^{4}}+\frac{h_{ij}\dot{N}^{i}\dot{N}^{j}}{N^{4}}+\frac{8N^{i}\dot{N}\partial_{i}N}{N^{4}}-\frac{2h_{ij}N^{k}\dot{N}^{j}\partial_{k}N^{i}}{N^{4}}\nonumber \\
 & -\frac{4\dot{N}^{i}\partial_{i}N}{N^{3}}-\frac{4N^{i}N^{j}\partial_{i}N\partial_{j}N}{N^{4}}-\frac{4h^{ij}K_{jk}\partial_{i}N^{k}}{N}+\frac{h_{ij}N^{k}N^{l}\partial_{k}N^{i}\partial_{l}N^{j}}{N^{4}}\nonumber \\
 & +\frac{4N^{i}\partial_{i}N^{j}\partial_{j}N}{N^{3}}-\frac{3\partial_{i}N^{j}\partial_{j}N^{i}}{N^{2}}+h^{ij}h^{kl}h^{mn}\partial_{k}h_{im}\partial_{n}h_{jl}-\frac{2h_{ij}h^{kl}\partial_{k}N^{i}\partial_{l}N^{j}}{N^{2}},
\end{align}
\begin{align}
Q_{\mu}Q^{\mu}= & -4K^{2}-\frac{8K\dot{N}}{N^{2}}-\frac{4\dot{N}^{2}}{N^{4}}+\frac{8KN^{i}\partial_{i}N}{N^{2}}+\frac{8N^{i}\dot{N}\partial_{i}N}{N^{4}}-\frac{8K\partial_{i}N^{i}}{N}\nonumber \\
 & -\frac{8\dot{N}\partial_{i}N^{i}}{N^{3}}+\frac{4h^{ij}h^{kl}\partial_{i}N\partial_{j}h_{kl}}{N}+\frac{4h^{ij}\partial_{i}N\partial_{j}N}{N^{2}}-\frac{4N^{i}N^{j}\partial_{i}N\partial_{j}N}{N^{4}}\nonumber \\
 & +\frac{8N^{i}\partial_{i}N\partial_{j}N^{j}}{N^{3}}-\frac{4\partial_{i}N^{i}\partial_{j}N^{j}}{N^{2}}+h^{ij}h^{kl}h^{mn}\partial_{m}h_{ij}\partial_{n}h_{kl},
\end{align}
\begin{align}
q_{\mu}q^{\mu}= & -\frac{4\dot{N}^{2}}{N^{4}}+\frac{h_{ij}\dot{N}^{i}\dot{N}^{j}}{N^{4}}+\frac{8N^{i}\dot{N}\partial_{i}N}{N^{4}}-\frac{4\dot{N}\partial_{i}N^{i}}{N^{3}}-\frac{2h_{ij}N^{k}\dot{N}^{j}\partial_{k}N^{i}}{N^{4}}\nonumber \\
 & -\frac{2h^{ij}\dot{N}^{k}\partial_{j}h_{ki}}{N^{2}}+\frac{2h^{ij}N^{k}\partial_{k}N^{l}\partial_{j}h_{li}}{N^{2}}+h^{ij}h^{kl}h^{mn}\partial_{l}h_{nk}\partial_{j}h_{im}\nonumber \\
 & -\frac{4N^{i}N^{j}\partial_{i}N\partial_{j}N}{N^{4}}+\frac{4N^{i}\partial_{i}N\partial_{j}N^{j}}{N^{3}}+\frac{h_{ij}N^{k}N^{l}\partial_{k}N^{i}\partial_{l}N^{j}}{N^{4}}-\frac{\partial_{i}N^{i}\partial_{j}N^{j}}{N^{2}},
\end{align}
\begin{align}
Q_{\mu}q^{\mu}= & -\frac{4K\dot{N}}{N^{2}}-\frac{4\dot{N}^{2}}{N^{4}}-\frac{h^{ij}\dot{N}^{k}\partial_{k}h_{ij}}{N^{2}}+\frac{4KN^{i}\partial_{i}N}{N^{2}}+\frac{8N^{i}\dot{N}\partial_{i}N}{N^{4}}-\frac{2K\partial_{i}N^{i}}{N}-\frac{6\dot{N}\partial_{i}N^{i}}{N^{3}}\nonumber \\
 & +\frac{h^{ij}N^{k}\partial_{k}N^{l}\partial_{l}h_{ij}}{N^{2}}-\frac{2\dot{N}^{i}\partial_{i}N}{N^{3}}-\frac{4N^{i}N^{j}\partial_{i}N\partial_{j}N}{N^{4}}+\frac{6N^{i}\partial_{i}N\partial_{j}N^{j}}{N^{3}}\nonumber \\
 & +h^{ij}h^{kl}h^{mn}\partial_{l}h_{nk}\partial_{m}h_{ij}+\frac{2N^{i}\partial_{i}N^{j}\partial_{j}N}{N^{3}}-\frac{2\partial_{i}N^{i}\partial_{j}N^{j}}{N^{2}}+\frac{2h^{ij}h^{kl}\partial_{i}N\partial_{l}h_{jk}}{N}.
\end{align}

\section{Primary constraints in various cases\label{app:pri_cons}}

The most general conjugate momenta are given in (\ref{pi_N})-(\ref{pi^ij}).
In this appendix, we give the explicit expressions for the primary
constraints in various cases.

\subsection{Case II}

In this case, the momenta become 
\begin{align}
\pi^{ij}= & \sqrt{h}\left[-4c_{1}K^{ij}-4c_{3}Kh^{ij}-4\left(2c_{1}+c_{2}\right)\frac{\partial_{k}N^{i}}{2N}h^{kj}\right],\\
\pi_{N}= & 4\sqrt{h}\left(c_{1}+c_{2}\right)\frac{\partial_{i}N^{i}}{N^{2}},\\
\pi_{i}\coloneqq & \sqrt{h}\bigg[-2\left(c_{1}+c_{3}\right)\frac{h_{ki}N^{j}\partial_{j}N^{k}}{N^{3}}-2\left(c_{3}-c_{1}-c_{2}\right)\frac{h^{kj}Q_{jik}}{N}\\
 & +2c_{3}\frac{h^{kj}Q_{ikj}}{N}-4\left(c_{2}-c_{3}\right)\frac{\partial_{i}N}{N^{2}}+2\left(c_{1}+c_{3}\right)\frac{h_{ij}\dot{N}^{j}}{N^{3}}\bigg].
\end{align}
We can see that $\dot{N}$ cannot be solved, which generates one primary
constraint
\begin{equation}
\phi_{N}\coloneqq\pi_{N}+4\left(c_{1}+c_{2}\right)\sqrt{h}\frac{\partial_{i}N^{i}}{N^{2}}\approx0.
\end{equation}

\subsection{Case III}

In this case
\begin{align}
\pi^{ij}\coloneqq & 4\sqrt{h}\left[c_{3}\left(3K^{ij}-Kh^{ij}\right)-\left(-6c_{3}+c_{2}\right)\frac{\partial_{k}N^{i}}{2N}h^{kj}\right],\nonumber \\
\pi_{N}\coloneqq & \sqrt{h}\left[-8\left(c_{2}-4c_{3}+c_{4}\right)\frac{\dot{N}}{N^{3}}-4\left(c_{4}-c_{3}\right)\frac{\partial_{i}N^{i}}{N^{2}}+8\left(c_{2}-4c_{3}+c_{4}\right)\frac{N^{i}\partial_{i}N}{N^{3}}\right],\nonumber \\
\pi_{i}\coloneqq & \sqrt{h}\bigg\{-2\left(-6c_{3}+c_{2}+c_{4}\right)\frac{h_{ki}N^{j}\partial_{j}N^{k}}{N^{3}}-2c_{4}\frac{h^{kj}Q_{jik}}{N}-\frac{c_{5}h^{kj}Q_{ikj}}{N}\nonumber \\
 & -4\left(c_{2}-c_{3}\right)\frac{\partial_{i}N}{N^{2}}+2\left(-6c_{3}+c_{2}+c_{4}\right)\frac{h_{ij}\dot{N}^{j}}{N^{3}}\bigg\}.
\end{align}
We can see that we cannot solve the trace part of $K_{ij}$, which
yields one primary constraint
\begin{equation}
\phi\coloneqq\pi+4\sqrt{h}\left(-6c_{3}+c_{2}\right)\frac{\partial_{i}N^{i}}{2N}\approx0.
\end{equation}

\subsection{Case IV}

In this case
\begin{align}
\pi^{ij}\coloneqq & \sqrt{h}\left[4c_{3}\left(3K^{ij}-Kh^{ij}\right)-4\left(-6c_{3}+c_{2}\right)\frac{\partial_{k}N^{i}}{2N}h^{kj}\right],\\
\pi_{N}\coloneqq & -4\sqrt{h}\left(3c_{3}-c_{2}\right)\frac{\partial_{i}N^{i}}{N^{2}},\\
\pi_{i}\coloneqq & \sqrt{h}\bigg[4c_{3}\frac{h_{ki}N^{j}\partial_{j}N^{k}}{N^{3}}-2\left(4c_{3}-c_{2}\right)\frac{h^{kj}Q_{jik}}{N}-\frac{c_{5}h^{kj}Q_{ikj}}{N}\\
 & -4\left(c_{2}-c_{3}\right)\frac{\partial_{i}N}{N^{2}}-4c_{3}\frac{h_{ij}\dot{N}^{j}}{N^{3}}\bigg].
\end{align}
We can see that $\dot{N}$ and $K$ cannot be solved, which yields
two primary constraints
\begin{align}
\phi\coloneqq & \pi+4\sqrt{h}\left(-6c_{3}+c_{2}\right)\frac{\partial_{i}N^{i}}{2N}\approx0,\\
\phi_{N}\coloneqq & \pi_{N}+4\sqrt{h}\left(3c_{3}-c_{2}\right)\frac{\partial_{i}N^{i}}{N^{2}}\approx0.
\end{align}

\subsection{Case V }

In this case
\begin{align}
\pi^{ij}\coloneqq & \sqrt{h}\bigg[-4c_{1}K^{ij}-4c_{3}Kh^{ij}-4\left(2c_{3}+c_{5}\right)\frac{\dot{N}}{2N^{2}}h^{ij}-4\left(2c_{3}+c_{5}\right)\frac{\partial_{i}N^{i}}{2N}h^{ij}\nonumber \\
 & +4\left(2c_{3}+c_{5}\right)\frac{N^{k}\partial_{k}N}{2N^{2}}h^{ij}-4\left(2c_{1}+c_{2}\right)\frac{\partial_{k}N^{i}}{2N}h^{kj}\bigg],
\end{align}
\begin{align}
\pi_{N}\coloneqq & \sqrt{h}\bigg\{-8\left(-c_{1}+c_{3}+c_{5}\right)\frac{\dot{N}}{N^{3}}-4\left(2c_{3}+c_{5}\right)\frac{K}{N}-2\left(4c_{3}-4c_{1}-2c_{2}+3c_{5}\right)\frac{\partial_{i}N^{i}}{N^{2}}\nonumber \\
 & +8\left(-c_{1}+c_{3}+c_{5}\right)\frac{N^{i}\partial_{i}N}{N^{3}}\bigg\},
\end{align}
\begin{align*}
\pi_{i}\coloneqq & \sqrt{h}\left[2\left(2c_{1}+c_{2}\right)\frac{h^{kj}Q_{jik}}{N}-\frac{c_{5}h^{kj}Q_{ikj}}{N}-2\left(2c_{2}+c_{5}\right)\frac{\partial_{i}N}{N^{2}}\right].
\end{align*}
In this case, $\dot{N}^{i}$ cannot be solved, which yields three
primary constraints
\begin{equation}
\phi_{i}\coloneqq\pi_{i}-\sqrt{h}\left[2\left(2c_{1}+c_{2}\right)\frac{h^{kj}Q_{jik}}{N}-\frac{c_{5}h^{kj}Q_{ikj}}{N}-2\left(2c_{2}+c_{5}\right)\frac{\partial_{i}N}{N^{2}}\right]\approx0.
\end{equation}

\subsection{Case VI}

In this case
\begin{align}
\pi^{ij}\coloneqq & \sqrt{h}\left[4c_{1}\left(-K^{ij}+Kh^{ij}\right)-4\left(2c_{1}+c_{2}\right)\frac{\partial_{k}N^{i}}{2N}h^{kj}\right],\\
\pi_{N}\coloneqq & 4\sqrt{h}\left(c_{1}+c_{2}\right)\frac{\partial_{i}N^{i}}{N^{2}},\\
\pi_{i}\coloneqq & \sqrt{h}\left[2\left(2c_{1}+c_{2}\right)\frac{h^{kj}Q_{jik}}{N}-2c_{1}\frac{h^{kj}Q_{ikj}}{N}-4\left(c_{2}+c_{1}\right)\frac{\partial_{i}N}{N^{2}}\right].
\end{align}
We can see that $\dot{N}^{i}$ and $\dot{N}$ cannot be solved, which
yields four primary constraints
\begin{align}
\phi_{N} & \coloneqq\pi_{N}-4\left(c_{1}+c_{2}\right)\sqrt{h}\frac{\partial_{i}N^{i}}{N^{2}}\approx0,\\
\phi_{i} & \coloneqq\pi_{i}-\sqrt{h}\left[2\left(2c_{1}+c_{2}\right)\frac{h^{kj}Q_{jik}}{N}-2c_{1}\frac{h^{kj}Q_{ikj}}{N}-4\left(c_{2}+c_{1}\right)\frac{\partial_{i}N}{N^{2}}\right]\approx0.
\end{align}

\subsection{Case VII}

In this case,

\begin{align}
\pi^{ij}\coloneqq & \sqrt{h}\left[4c_{3}\left(3K^{ij}-Kh^{ij}\right)-4\left(-6c_{3}+c_{2}\right)\frac{\partial_{k}N^{i}}{2N}h^{kj}\right],\\
\pi_{N}\coloneqq & \sqrt{h}\left[-16c_{3}\frac{\dot{N}}{N^{3}}-4\left(5c_{3}-c_{2}\right)\frac{\partial_{i}N^{i}}{N^{2}}+16c_{3}\frac{N^{i}\partial_{i}N}{N^{3}}\right],\\
\pi_{i}\coloneqq & \sqrt{h}\left[2\left(-6c_{3}+c_{2}\right)\frac{h^{kj}Q_{jik}}{N}+2c_{3}\frac{h^{kj}Q_{ikj}}{N}-4\left(c_{2}-c_{3}\right)\frac{\partial_{i}N}{N^{2}}\right].
\end{align}
We can see that $\dot{N}^{i}$ and the trace part of $K_{ij}$ cannot
be solved, which yields four primary constraints
\begin{align}
\phi\coloneqq & \frac{\pi}{\sqrt{h}}+4\left(-6c_{3}+c_{2}\right)\frac{\partial_{i}N^{i}}{2N},\\
\phi_{i}\coloneqq & \frac{\pi_{i}}{\sqrt{h}}-2\left(-6c_{3}+c_{2}\right)\frac{h^{kj}Q_{jik}}{N}-2c_{3}\frac{h^{kj}Q_{ikj}}{N}+4\left(c_{2}-c_{3}\right)\frac{\partial_{i}N}{N^{2}}.
\end{align}

We do not show the explicit expressions of the primary constraints
of cases VIII--XIII (i.e., cases in Table \ref{tab:cases_unphy}).
These cases have the same condition $c_{1}=0$, which means there
are no gravitational waves. Therefore, we do not consider these cases
in the present paper.

\section{Quadratic action for the scalar perturbations \label{app:S2conc} }

The full expression of the quadratic action for the scalar perturbations
of the QSN theory is given by
\begin{align}
S_{2,\mathrm{s}}= & \int\mathrm{d}t\mathrm{d}^{3}x\Big[-6\left(c_{1}+3c_{3}\right)a\dot{a}^{2}A^{2}+12\left(2c_{3}+c_{5}\right)a^{2}\dot{a}A\dot{A}-4\left(c_{1}+c_{2}+c_{3}+c_{4}+c_{5}\right)a^{3}\dot{A}^{2}\nonumber \\
 & +24\left(c_{1}+3c_{3}\right)a^{2}\dot{a}A\dot{\zeta}-12\left(2c_{3}+c_{5}\right)a^{3}\dot{A}\dot{\zeta}-12\left(c_{1}+3c_{3}\right)a^{3}\dot{\zeta}^{2}+\frac{1}{4}a^{3}\dot{\phi}^{2}_{0}A^{2}+\frac{1}{2}a^{3}VA^{2}+\tfrac{1}{2}a^{3}\dot{\delta\phi}^{2}\nonumber \\
 & +12\left(c^{\prime}_{1}+3c^{\prime}_{3}\right)a\dot{a}^{2}A\delta\phi-12\left(2c^{\prime}_{3}+c^{\prime}_{5}\right)a^{2}\dot{a}\dot{A}\delta\phi-24\left(c^{\prime}_{1}+3c^{\prime}_{3}\right)a^{2}\dot{a}\dot{\zeta}\delta\phi\nonumber \\
 & +a^{3}V_{\phi}A\delta\phi+\tfrac{1}{2}a^{3}\delta\phi^{2}V_{\phi\phi}-6\left(c^{\prime\prime}_{1}+3c^{\prime\prime}_{3}\right)a\dot{a}^{2}\delta\phi^{2}+36\left(c_{1}+3c_{3}\right)a\dot{a}^{2}A\zeta-36\left(2c_{3}+c_{5}\right)a^{2}\dot{a}\dot{A}\zeta\nonumber \\
 & -72\left(c_{1}+3c_{3}\right)a^{2}\dot{a}\dot{\zeta}\zeta-\frac{3}{2}a^{3}\dot{\phi}^{2}_{0}A\zeta+3a^{3}VA\zeta-36\left(c^{\prime}_{1}+3c^{\prime}_{3}\right)a\dot{a}^{2}\delta\phi\zeta+3a^{3}V_{\phi}\delta\phi\zeta-54\left(c_{1}+3c_{3}\right)a\dot{a}^{2}\zeta^{2}\nonumber \\
 & +\frac{9}{4}a^{3}\dot{\phi}^{2}_{0}\zeta^{2}+\frac{9}{2}a^{3}V\zeta^{2}-4\left(c_{1}+c_{3}\right)aA\Box A-2\left(2c_{2}+12c_{3}+7c_{5}\right)a\dot{a}B\Box A-4\left(6c_{3}+c_{5}\right)a\zeta\Box A\nonumber \\
 & +2\left(2c_{2}+3c_{5}\right)a\dot{a}A\Box B-\left(2c_{1}+c_{2}+c_{4}\right)aB\dot{a}^{2}\Box B-2\left(2c_{4}+c_{5}\right)a^{2}\dot{A}\Box B\nonumber \\
 & -2\left(2c_{2}+3c_{5}\right)a^{2}\dot{\zeta}\Box B-2\left(2c^{\prime}_{2}+3c^{\prime}_{5}\right)a\dot{a}\delta\phi\Box B-4\left(6c_{1}+3c_{2}+18c_{3}+c_{4}+6c_{5}\right)a\dot{a}\zeta\Box B\nonumber \\
 & -4\left(3c_{1}+c_{2}+9c_{3}+c_{4}+3c_{5}\right)a\zeta\Box\zeta+2\left(2c_{2}+c_{5}\right)a^{2}A\Box\dot{B}+2\left(2c_{1}+c_{2}+c_{4}\right)a^{2}\dot{a}B\Box\dot{B}\nonumber \\
 & +2\left(2c_{4}+3c_{5}\right)a^{2}\zeta\Box\dot{B}+a^{2}\dot{\phi}_{0}\delta\phi\Box B+\frac{1}{2}a\delta\phi\Box\delta\phi-\left(2c_{1}+c_{2}+c_{4}\right)a\left(\Box B\right)^{2}\nonumber \\
 & -\frac{4}{3}c_{5}a\Box A\Box E-\frac{4}{3}c_{4}a\dot{a}\Box B\Box E-\frac{4}{3}\left(2c_{2}+2c_{4}+3c_{5}\right)a\Box E\Box\zeta\nonumber \\
 & +\left(c_{1}+c_{2}\right)a\square\partial^{i}E\square\partial_{i}E+\frac{1}{9}\left(3c_{1}+5c_{2}-4c_{4}\right)a\Box E\Box\Box E-a^{3}A\dot{\phi}_{0}\delta\dot{\phi}\nonumber \\
 & +3a^{3}\dot{\phi}_{0}\zeta\delta\dot{\phi}+\frac{4}{3}c_{4}a^{2}\Box E\Box\dot{B}-\left(2c_{1}+c_{2}+c_{4}\right)a^{3}\dot{B}\Box\dot{B}-\frac{4}{3}c_{2}a^{2}\Box B\Box\dot{E}-\frac{2}{3}c_{1}a^{3}\Box\dot{E}\Box\dot{E}\Big].
\end{align}
Note in deriving the above, no integration by parts has been performed.
By submitting the conditions of coefficients in various cases, we
can get the quadratic actions in Secs. \ref{subsec:pert_case2}--\ref{subsec:pert_case6}.

\providecommand{\href}[2]{#2}\begingroup\raggedright\endgroup

\end{document}